\documentclass[natbib]{svjour3}      

\smartqed  

\usepackage{graphicx}
\usepackage{float}
\usepackage{amsfonts}
\usepackage{amsmath}
\usepackage{url}
\usepackage{blindtext}
\DeclareUrlCommand\url{\def\UrlLeft{<}\def\UrlRight{>}
\urlstyle{tt}}

\usepackage{color}
\newcommand{\real}{\mathbb{R}}
\newcommand{\num}{num\'{e}raire}
\newcommand{\nume}{num\'{e}raire }

\begin{document}

\title{Price Dynamics Via Expectations,\\ and the Role of Money
Therein
\thanks{We acknowledge funding by the EU, Horizon 2020 programme, DOLFINS
project (No. 640772)}}

\author{Gesine A. Steudle\authormark{1}  \and 
 				Saini Yang \authormark{2}                      \and
        Carlo C. Jaeger\authormark{2}\authormark{1}\corresponding
}

\institute{\authormark{1}Global Climate Forum  \\[2mm]
					 \authormark{2}Beijing Normal University \\[2mm]					 
           \corresponding \email{carlo.jaeger@globalclimateforum.org}
           }

\date{December 1, 2016}

\issue{3/2016}

\titlepage

\maketitle

\begin{abstract}
Beyond its obvious macro-economic relevance, fiat money has important
micro-economic implications. They matter for addressing No.\ 8 in Smale's
``Mathematical Problems for the Next Century'' (\cite{smale1998}): extend the
mathematical model of general equilibrium theory to include price adjustments.
In the canonical Arrow-Debreu framework, equilibrium prices are set by a
fictitious auctioneer.
Removing that fiction raises the question of how prices are set and adjusted by
decentralised actors with incomplete information. We investigate this question
through a very basic model where a unique factor of production, labour, produces
a single consumption good, called jelly for brevity. The point of the model is
to study a price dynamics based on the firm's expectations about jelly demand
and labour supply. The system tends towards economic equilibrium, however,
depending on the initial conditions it might not get there. In different
model versions, different kinds of money are introduced. Compared to the case
of no money, the introduction of money as a store of value facilitates the
system reaching economic equilibrium. If money is introduced as a third
commodity, i.e. there is also a demand for money, the system dynamics in general
becomes more complex.
\end{abstract}

%=======================================================================================

\section{Introduction}

In the Arrow-Debreu framework (\cite{arrow+debreu1954}), an auctioneer
establishes equilibrium prices. In intertemporal models, these prices may
display all sorts of dynamics, always brought about by the auctioneer.
Unfortunately, the auctioneer is an auxiliary entity not to be found in actual
economies. However, attempts to remove the auctioneer from economic models have
consistently run in the unsolved question of how to understand and model
non-equilibrium price dynamics.

In a non-equilibrium situation, usually it is assumed that prices somehow
adjust according to excess demand ($\dot{p}\propto \xi(p)$, see e.g.
\cite{saari1995}). But making this assumption, still the system dynamics can
become arbitrarily complex, and it is not clear whether a stable equilibrium
will be obtained, a fact which is known as the Sonnenschein-Mantel-Debreu (SMD)
Theorem (\cite{sonnenschein1972}, \cite{mantel1974}, \cite{debreu1974}).
An overview of the problems arising from the SMD Theorem is given in
\cite{rizvi2006}.

In the real world, when there is what is usually called a free market, the
dynamics of prices result from the behaviour of a large number of different
actors. Theoretically, such a system can be simulated using an agent-based model
with many agents following their decision rules. An important attempt of 
understanding the dynamics of general equilibrium by agent-based modelling can
be found in \cite{gintis2007}. In such models, agents can have incomplete
information and take decisions based on expectations (e.g. \cite{an+al2007}).
For modelling price dynamics it is not too easy a challenge to model the
behaviour of the agents in detail (since it is also not known in detail how real
agents behave), and thus usually very rough assumptions are made, supposing that
agents try to optimise their own utility. On the other hand, in many cases the
general-equilibrium picture of clearing markets seems to be a good
representation of what is happening in the real world. So the agent-based model
allowing us to abandon the auctioneer should still be able to deliver these
equilibrium results. The additional value of such an agent-based model is then
supposed to be a better understanding of (real world) situations where an
economic equilibrium is not obtained or disturbed (\cite{farmer+foley2009}).

In this paper, the approach for getting rid of the auctioneer and modelling an
out-of equilibrium price-dynamics is the following: A price setting entity
observes excess demand at the current price and updates the price for the next
period accordingly.
Let's assume that the price setting party seeks economic equilibrium but does
not have complete information (because in that case it would be the auctioneer).
Observation of some previous price and excess demand pairs $(p,\xi(p))$ does not
suffice to determine the next price. Additional assumptions have to be made,
and since there is no further insight into the future decisions of other
actors, these additional assumptions of the price setter have to be grounded in
expectations about the aggregate supply and demand. But this means that the
present state of the system is influenced by expectations about the future.
Linking the present state with expectations of the future, however, is what
Keynes identified as one of the essential characteristics of money:
``Or, perhaps, we might make our line of division between the theory of
stationary equilibrium and the theory of shifting equilibrium -- meaning by the
latter the theory of a system in which changing views about the future are
capable of influencing the present situation. For the importance of money
essentially flows from its being a link between the present and the future.''
(\cite{keynes1936}, Chapter 21) In this work, we investigate how the
existence and amount of money in our very basic model influences the system's
dynamics. Although here the availability of money is not taken into account when
expectations are formed, we see that it limits the possibilities of the system
to evolve over time. The amount of money thus influences which state the system
converges to, i.e. whether starting out of equilibrium an economic equilibrium
is obtained eventually.

\begin{figure}
\begin{center}
\includegraphics[width=8.5cm]{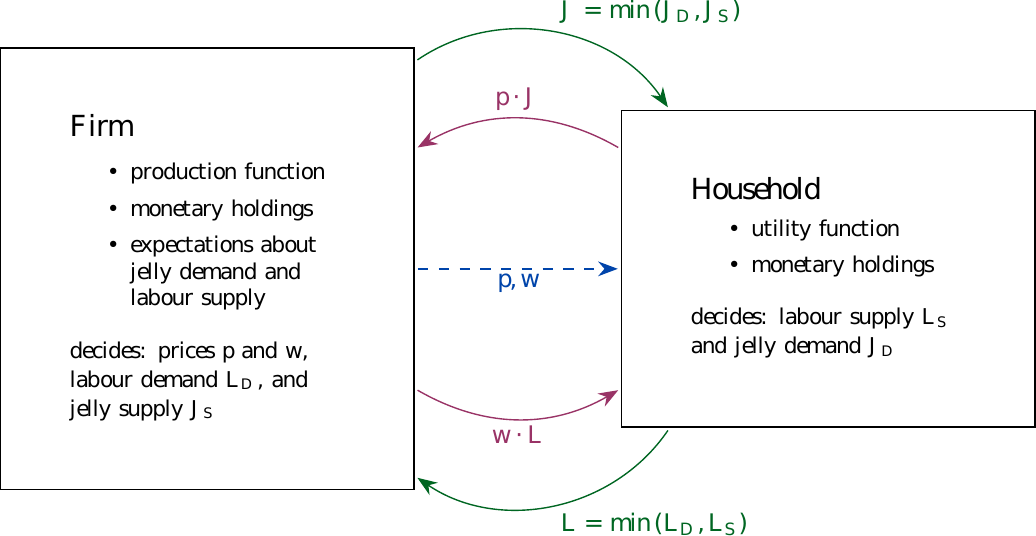}
\caption{Model overview \label{F:modelOverview}}
\end{center}
\end{figure}

For understanding some very fundamental principles, we intended to use a model
as simple as possible: 
\begin{itemize}
  \item There are two aggregate agents and two goods. Good prices are set by one
  of the agents, with the other one reacting to the proposed price. The agents interact
  repeatedly. In our model, the agents are an aggregate firm that produces a
  consumption good (for brevity called jelly) from labour, and an aggregate
  household that supplies labour and consumes jelly.
  \item The agents have preferences/decision rules that determine their
  supply and demand of the different goods. The aggregate firm optimises its
  profit and the aggregate household its utility to determine supply and demand.
  \item The price setting agent has expectations about the relation of
  prices and aggregate supply/demand.	In our model, prices are set by the firm.
  \item The price setting agent updates his expectations using observations
  about the actual excess demand at the prices that have been set before.
\end{itemize}
Figure \ref{F:modelOverview} gives an overview of the agents and their
interactions. There are three versions of the model which differ in whether and
how money is part of the economic system.

The remainder of the paper is organised as follows: In Section \ref{S:agents}
the agents are described in detail, and it is explained how transactions between
agents take place. The three different model versions are discriminated. In
Section \ref{S:dynamics} the dynamics of the system is discussed, and it is analysed
which are the possible states (equilibria) it converges to. Section
\ref{S:simulations} shows some simulations illustrating the findings of Section
\ref{S:dynamics}. Sections \ref{S:dynamics} and \ref{S:simulations} only deal
with the first two versions of the model in which -- although there might be
something like ``auxiliary money'' -- labour serves as \num. In Section
\ref{S:realMoney} the third version of in the model is introduced. Here money
has the role of a third commodity (and is the \num). Section \ref{S:conclusions}
concludes the paper.

%=======================================================================================

\section{Agents, Markets, and Money}
\label{S:agents}

\subsection{Household}
\label{S:household}

The objective of the household is to maximise its utility subject to its
budget constraint. Here, we have chosen a standard Cobb-Douglas utility function
$U(L,J)$ depending on labour $L$ and jelly consumption $J$:
\begin{equation}
U(L,J)=(L_f-L)^\alpha \cdot J^\beta \label{E:utility}
\end{equation}
$L_f$ is the maximum amount of labor force available. As usual, households are
assumed to maximise $L_f - L$, interpreted as leisure time, and consumption $J$
subject to a budget constraint. In the versions of the model we focus on in this
paper\footnote{These are the two of three versions, for a discrimination see
Section \ref{S:money}. The household's utility optimisation for the third
version is given in Section \ref{S:household2}}, the budget constraint of the
household is given by
\begin{equation}
p\cdot J = w\cdot L. \label{E:budgetConstraint}
\end{equation}
$p$ is the price of jelly, $w$ the wage. Equation (\ref{E:budgetConstraint})
means that the household plans to spend its total income to buy jelly.

\begin{figure}
\begin{center}
\includegraphics[width=4.6cm]{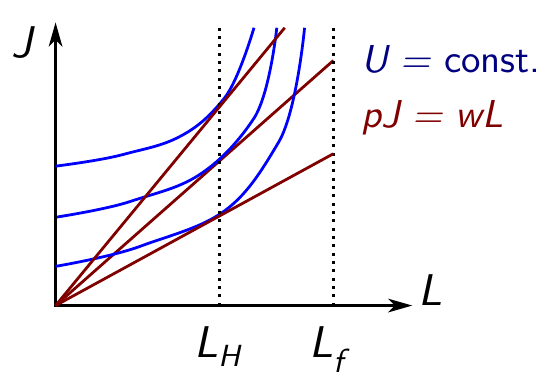}
\caption{Budget lines, $J=\frac{w}{p}L$, for different $w$-$p$-ratios (red)
and indifference curves, $J=\sigma_c(L)$, for different constants $c$ (blue) in
the $L$-$J$-plane. The maxima of $U$ subject to $p\cdot J=w\cdot L$ all lie on
the vertical at $L=L_H$. \label{F:indifference}}
\end{center}
\end{figure}

Maximising the utility given in (\ref{E:utility}) subject to the budget
constraint (\ref{E:budgetConstraint}) yields a utility maximum at
\begin{eqnarray}
L_H &=& \frac{\beta}{\alpha+\beta} \cdot L_f \label{E:plannedLS}\\
J_H &=& \frac{w}{p} \cdot \frac{\beta}{\alpha+\beta} \cdot L_f
\label{E:plannedJD}
\end{eqnarray}
The Cobb-Douglas form of (\ref{E:utility}) together with constraint
(\ref{E:budgetConstraint}) implies that $L_H$ only depends on $\alpha, \beta,
L_f$ and not on the wage-to-price ratio $w/p$. Indifference curves have the form
\begin{equation}
\sigma_c (L)  =  \frac{c}{ (L_f-L)^{\frac{\alpha}{\beta}}} \label{E:indCurve}
\end{equation}
In Figure \ref{F:indifference}, utility maxima are shown for different
wage-to-price ratios $\frac{w}{p}$. 

\subsection{Firm}

\subsubsection{Profit Optimisation}
\label{S:profitOpt}

In a situation of perfect competition of many firms, the aggregate firm must
realise zero profits (\cite{vonneumann1945}). Thus, maximisation of actual
profits by individual firms results in minimisation of absolute profits by the
aggregate firm. Therefore, the aggregate firm minimises the absolute value of
the difference of revenues and expenditures:
\begin{equation}
	\min \vert pJ - wL \vert \label{E:profitOpt}
\end{equation}

In most of what is discussed in this paper, labour serves as \num, i.e. $w=1$.
The production function is assumed to be $J = \rho(L) = L^\gamma$ (with
$0<\gamma<1$). Let's assume that the firm's demand expectations can be
represented by a function $\phi$ (that means at a jelly price of $\phi(J)$ the
firm expects the jelly demand to be $J$). The optimisation problem becomes
\begin{eqnarray}
	\min_{L}& &\vert pJ - L \vert\\
	\text{s.t.}& &J = \rho(L) =  L^\gamma \label{E:productionFn}\\
	& & p = \phi(J)\label{E:price}\\
	& & 0 \leq L \leq L_f,\label{E:labourF}
\end{eqnarray}
that means it can be written as 
\begin{equation}
	\min_{0 \leq L \leq L_f} \vert \rho(L) \cdot \phi(\rho(L)) - L \vert.
	\label{E:profitOptL}
\end{equation}

In the model, the firm optimizes its expected profit to determine prices,
planned labour demand, and planned jelly supply. The firm's planned labour
demand is the argument of the profit optimisation given in (\ref{E:profitOptL}).
Whether it is also the actual labour demand depends on whether the firm can
afford to employ as much labour as it would like to, as explained in Section
\ref{S:evoRules}. An extension of the firm's optimisation for the case that
money serves as \nume and thus wages have to be set by the firm as well is given
in Section \ref{S:profitOpt2}.

\subsubsection{Expectations and Learning}
\label{S:expAndLearn}

\begin{figure}
\begin{center}
\includegraphics[width=13.4cm]{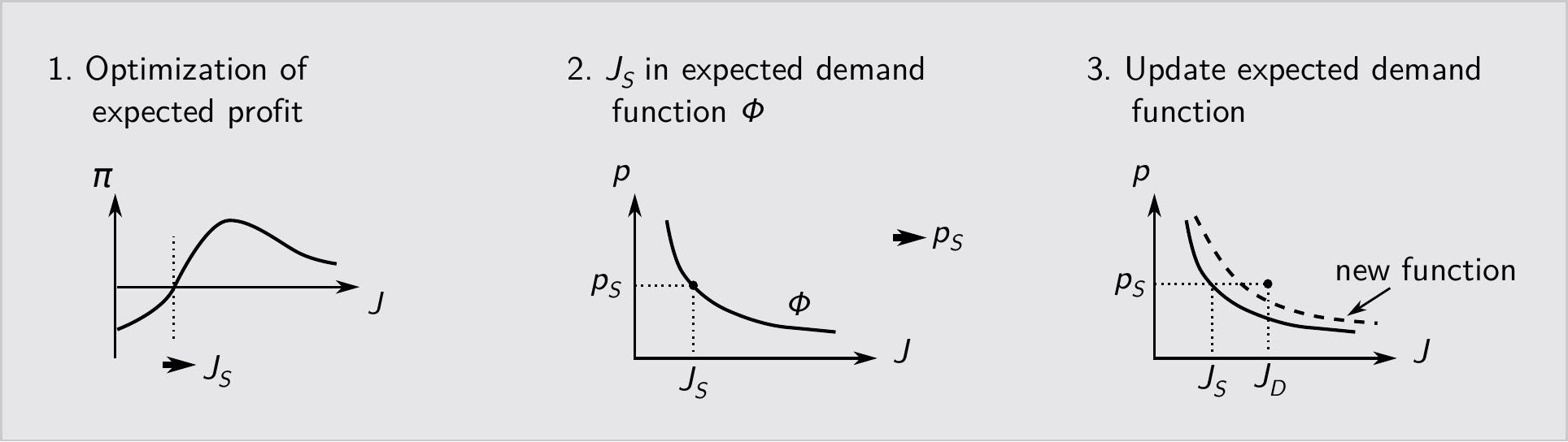}
\caption{Scheme of firm's learning algorithm: 1. Based on its demand
expectations, the firm optimises its expected profit. This determines how much
jelly $J_S$ the firm wants to supply. 2. To set the price, the firm evaluates its expected
demand function $\phi$ at $J=J_S$. The outcome $p_S=\phi(J_S)$ is set as jelly
price. 3. At price $p_S$, the jelly demand $J_D$ might be different from $J_S$.
The firm uses this information to update its expected demand function. In
the next time step, the firm uses this new expected demand function. 
\label{F:learningFirm}}
\end{center}
\end{figure}

The firm's expectations have standard textbook form. The firm assumes a falling
demand curve for jelly (and, for the case that labour is not the \num, a rising
supply curve for labour).
The updates of its expectation functions are based on observations of the
household's jelly demand (and, in the third version of the model, labour
supply). At each time step, information about the actual supply and demand at
the current price is obtained by the firm and this information is used to update the
expectation function(s). However, it is assumed that the firm also has  a
certain unwillingness to change its expectations essentially at every time step.
A schematic overview of the algorithm used to update the expected labour
demand function is given in Figure \ref{F:learningFirm}.

The updating algorithm adjusts the parameters for the expected demand (and
supply) functions recursively. It approximates the last two observations. For
the functional forms of $\phi$ and the updating algorithm used in the below
simulations see the appendix.

\subsection{Markets and Money}
\label{S:money}

Having introduced the agents and their decision rules, the question arises
whether and how the agents can fulfill their consumption/production plans. For
producing/consuming they need to interact mutually which is -- as usual --
supposed to happen on markets. The fact that the model allows for
non-equilibrium states requires rules for what happens for non-clearing markets.

In an exchange economy in equilibrium, transactions in all markets can be
thought of as taking place instantaneously (although a temporal order may be
natural, as e.g.\ in the case that labour has to be hired in order to produce a
good sold in another market). All actors optimise their respective objective
functions and prices are set by the auctioneer in a way that all markets clear.
This means that all the agents' plans of how much of all goods to buy or sell
can be fulfilled.

On the other hand, in a non-equilibrium situation, unmet supply or demand in one
market can make it impossible for actors to follow their plans in other markets
like in the above production example: if a producer is not able to hire as much
labour as she has planned, she may not be able to produce as much as she
had planned and thus her supply in the production good market may differ from
her planned supply. This illustrates that for a non-equilibrium situation the
temporal order of transactions in the different markets matters. However, the
exchange of goods is usually thought of as an instantaneous act, otherwise the
value of the good changing owners first has somehow to be stored by the seller
until he buys the other good for this stored value. This means that value has to
be stored at least for a short time. It suggests the introduction of (something
like) money as a store of value for non-equilibrium situations.  

In particular, in our model there are two markets, a labour and a jelly market,
with the following features:
\begin{itemize}
  \item In both markets, if supply does not equal demand, the short side of the
	market prevails.
	\item In every time step, first transactions take place in the labor market and 
	later in the jelly market.
	\item In both markets, the buyer is limited by her budget, i.e. there is an
	upper limit of how much she can afford. If there is money, this upper limit is
	given by her cash balance. If there is no money, things become a bit trickier:
	for the household (buyer in the jelly market) the upper limit is given by the
	value of labour it sold at the last transaction in the labour market. For the
	firm (buyer in the labour market) the budget is given by the value of jelly it
	sold at the last transaction in the jelly market. 
\end{itemize}
Introducing money in the system makes the system more complex. For example,
it raises the questions of supply and demand for money, and if there is a demand
for money, this means that more expectations are introduced to the
model\footnote{The demand for money is based on the expectation that it will be
useful at some later point in time.}.
However, the fact that we need to store value - at least for the short time
between transactions in the two markets - suggests the introduction of some kind
of money. in this paper we focus on two versions of the model, one without money
and one with something like ``light'' money:
\begin{enumerate}
  \item First version (no money): There is no money, only jelly and labour of
  equal value can be exchanged (quasi-)instantaneously, that means in the
  implementation two transactions are executed one after another (thus the value
  of labour or jelly can be stored for a short period of time, i.e. from the
  point of time transactions take place in one market to the point they take
  place in the other market).\\
  \item Second version (money version 1): ``Money'' exists as a store of value,
  and in the beginning firm and household may have a certain amount of money at
  their disposal. However, for the household's utility optimisation money is
  irrelevant, as well as in the firm's profit optimisation. That means that
  there is no demand for money.
  Labour is still the \num, and the price of money equals one, i.e. one unit of
  money always has the same value as one unit of labour. Money is used to store
  value in units of labour. 
  \end{enumerate}
At the end of the paper, with a third model version we give an outlook of how to
modify the money ``light'' to make it become more similar to real money:
  \begin{enumerate}
	\item[3.] In the third version (money version 2) it is assumed that the
	household wishes to keep a fraction $s$ of its current income in cash (e.g. to cover
	unforeseen costs in the future). This is implemented by changing the boundary
	condition in the household's utility optimisation, as given in equation
	(\ref{E:budgetConstraintM}). This means that a demand for money is created.
	Money serves as \num, the price of labour may be different from one. The firm
	sets the wage based on its expectations about the labour supply in the same way
	as the jelly price is set based on expectations about the jelly demand (see
	Section \ref{S:expAndLearn}). Thus, in money version 2, the two-commodity model
	is transformed into a three-commodity model.\\
\end{enumerate}

In the following (Sections \ref{S:dynamics} and \ref{S:simulations}), we focus
on the first two cases. A discussion of money version 2 is given in Section
\ref{S:realMoney}.

%=======================================================================================

\section{System Dynamics}
\label{S:dynamics}

Often it is assumed that economies are usually in economic equilibrium (all
markets clear), and that this is a stable state. That means if they are pushed
out of it they will come back to it quickly. This common assumption, however, is
not part of general equilibrium theory (see e.g. \cite{sonnenschein1972}). So
when we look at the dynamics of our simple model we are particularly interested
in whether an initial out-of-equilibrium state of the system always converges to
an economic equilibrium. But firstly, we still have to define rules for how our
systems evolves over time which is done in the first part of this section. In
the second part it is discussed how the economic equilibria look like in our
model and the last part addresses the question just raised, that is whether the
systems obtains economic equilibrium eventually.

\subsection{Evolution Rules}
\label{S:evoRules}

If an aggregate household and firm interact along the lines indicated above, it
is useful to consider an iterated cycle involving the following steps:
\begin{enumerate}  
	\item \textit{Firm plans production}: The firm has expectations about the
		household's jelly demand and labour supply. By optimizing its expected profit
		the firm plans the jelly production and thus the labour demand, and sets the
		jelly price $p$ accordingly. However, the labour demand $L_D$ is limited by
		the firm's budget. In the version without money, the budget limitation is
		given by the revenues of the last jelly transaction, in the case that there is
		money by the firm's monetary holdings.%1
  \item \textit{Household plans consumption}: Optimising its utility function
  	subject to its budget constraint the household decides how much labour $L_S$
  	to offer and how much jelly it plans to consume given price $p$.%2
  \item \textit{Labour market transactions}: In the labour market, the short
  	side of the market prevails, i.e. $L_M=\min\{L_D,L_S\}$.%3
  \item \textit{Firm produces jelly}: Using $L_M$ the firm produces the actual
  	jelly supply $J_S$. If $L_M=L_D$, $J_S$ equals the planned supply but if
  	$L_M<L_D$, $J_S$ is less than the planned production.%4
  \item \textit{Household decides jelly demand}: If $L_M=L_S$ the household's
  	jelly demand $J_S$ equals the planned consumption, if $L_M<L_S$ the household
  	demands as much jelly as it can afford. If there is no money, this always
  	means $J_D=w/p \cdot L_M$, with money the jelly demand might be different
  	since the household can also use money from the last period to acquire
  	jelly.%5
  \item \textit{Jelly market transactions}: In the jelly market, the short
  	side of the market prevails, i.e. $J_M=\min \{J_D,J_S\}$.%6
  \item \textit{Expectations update}: Using the transaction information (actual
  	jelly demand $J_D$ at price $p$) the firm updates its jelly demand
  	expectation function.%7
\end{enumerate}

With money, the cycle is close to the intuition of everyday life in a
modern economy. Where supply and demand do not match, the difference will affect
the money holdings of the agents. If the firm can hire less labour than it
intended, it will remain with more money than it planned, if the household can
sell less labour than it intended, it will remain with less money than it
planned. The analogous pattern arises with sales of jelly.\footnote{Of course,
monetary holdings can also change when markets clear. The difference is that
if they do not clear, the monetary holdings must be affected.}

With money the monetary holdings are a constraint for what an agent can plan to
buy and also for what she may actually buy when plans cannot be realised.
Without money we set an analogous constraint. %in terms of the \nume (labour).
The labour sold by the household in step 3 then defines a budget constraint for step
5, and the jelly sold by the firm in step 5 defines a budget constraint for step
3 in the subsequent cycle.

It might be useful to point out that in this paper we use the term
\textit{budget constraint} for the budget constraint the household takes into
account when optimising its utility, given in equation
(\ref{E:budgetConstraint}). On the other hand, we use the term \textit{budget
limitations} for the limitations household and firm might experience when they
want to meet their demands but their monetary holdings, or for the case without
money their revenues from the last market interaction, do not suffice to do that
completely (see steps 1 and 5). Budget limitations are not taken into account in
the respective optimisations.

\subsection{Economic Equilibria}

The term equilibrium is used in different ways, for our purposes it is useful to
clearly differentiate between economic equilibria, i.e. states of the system
where supply equals demand in all markets, and dynamic equilibria, i.e. fixed
points of the dynamic system. Of course, economic equilibria can be fixed
points, and vice versa, but this is not necessarily the case. \\

\begin{figure}
\begin{center}
\includegraphics[width=4.6cm]{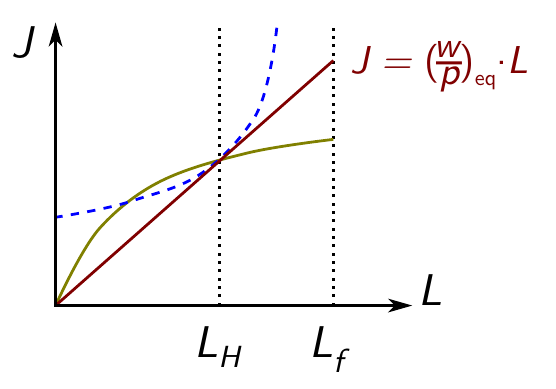}
\caption{Equilibrium wage-price-ratio $(\frac{w}{p})_\mathrm{eq}$.\label{F:equilibriumB}}
\end{center}
\end{figure}

The system being studied here has only one economic equilibrium. Labour
serves as \num, so $w=1$. The household's utility maximisation subject to its
budget constraint yields a maximum at
\begin{eqnarray}
L_H &=& \frac{\beta}{\alpha+\beta} \cdot L_f \label{E:LE}\\
J_H &=& \frac{1}{p} \cdot \frac{\beta}{\alpha+\beta} \cdot L_f \label{E:JE} 
\end{eqnarray}
(see Section \ref{S:household}). At the economic equilibrium the labour supply
$L_S = L_H$ and the jelly demand $J_D=J_H$. In the $L$-$J$-plane
$(L_H,J_H)$ is the point where the budget line $J=w/p\cdot L$ intersects with
the vertical at $L=L_H$.
For the firm, in the $L$-$J$-plane the planned production is represented by
the point of intersection of the budget line $J=w/p\cdot L$ and the production
function $J=L^\gamma$.
Economic equilibrium requires supply to equal demand in both markets, i.e.
for equilibrium the budget line has to intersect with the production
function at $L=L_H$ (see Figure \ref{F:equilibriumB}). Setting again $w=1$, thus
$J_H = (L_H)^\gamma$ determines the equilibrium price $p_E$:
\begin{equation}
 	\frac{1}{p_E}  = \left( \frac{\beta}{\alpha+\beta}
 	L_f \right)^{\gamma-1}. \label{E:wpE}
\end{equation}

\subsection{Dynamic Equilibria}
\label{S:fixedPoints}

Dynamic equilibria are the fixed points (steady states) of the system, i.e. the
points in state space that once the system has reached them it does not leave
them any more. However, what we are interested here is rather not the evolution
of all state variables (e.g. the parameters of the firm's expected demand
function which don't really have a ``real world'' equivalent) but only the
evolution of the variables that matter economically, which are supplies,
demands, and prices. We suppose that our economy has reached a fixed point if
these ``economic'' variables do not change any more over time.

\begin{definition}[Economic space, economic states]
Labour supply $L^{S}_t$ and demand $L^{D}_t$, jelly supply $J^{S}_t$ and demand
$J^{D}_t$, and the jelly price $p_t$ at time $t$ are state variables of the
system. The space spanned by these variables is called the economic space
$\mathcal{E} \subseteq \real^5_+$. Let $s_t = (L^{D}_t, L^{S}_t, J^{D}_t,
J^{S}_t, p_t, \ldots)$ be the state of the system at time $t$, i.e. the vector
of all state variables. Then the projection of the state $s_t$ on the economic
space, i.e. the vector $e_t = (L^{D}_t, L^{S}_t, J^{D}_t, J^{S}_t, p_t) \in
\real^5_+$ is called the economic state of the system at time $t$.
\end{definition}

The system under study is deterministic. This means that state space
trajectories do not intersect. To know that for time $t=0$ the system is at 
at a certain point $s_0$ in state space suffices for knowing the systems
behaviour for all $t>0$. Since some of the necessary information is lost
projecting states onto economic states, this is not true for trajectories in the
economic space\footnote{The system's evolution in the economic space can be
considered as the evolution of a a non-deterministic system with the source of
the uncertainty resulting from the lack of knowledge of some ``hidden
variables'' (like e.g. the parameters $z_t, \zeta_t$ of the firm's expected
demand function $\phi_t(J)=\frac{z_t}{J^{\zeta_t}}$).}.
As explained above, we are interested in the dynamic equilibria in the economic
space, and in whether they are economic equilibria as well. It turns out, that
while the economic equilibrium is also dynamic equilibrium, also other dynamic
equilibria can be found in which not all the markets clear.

\begin{proposition}[Types of fixed points]\label{P:types}
In the economic space of the system, there are two types of fixed points, the
economic equilibrium and one other type which will be called ``border
equilibrium''. In a border equilibrium, at least one market does not clear.
\end{proposition}

To understand this, in the following we consider how trajectories in the
economic space look like and how the agents' budget limitations restrict them.

\begin{definition}[Learning trajectory]
The state space learning trajectory $\{s_t\}$ of an initial state
$s_\text{in}=s_0$ is the sequence of states $\{s_t\}$ that are obtained after
$t$ time steps if the system's evolution rules are applied without any budget
limitations. The learning trajectory $\{l_t\}$ is the projection of $\{s_t\}$
onto the economic space.
\end{definition}

\begin{proposition}\label{P:convergence}
All learning trajectories converge to the the economic equilibrium $e_E$,
i.e. $\lim_{t\to \infty} l_t = e_E$ for all initial states $s_\text{in}$.
\end{proposition}

Proposition \ref{P:convergence} means that the firm's learning mechanism
directs the system towards the economic equilibrium. However, the learning
trajectory does not take into account budget limitations. At each time step,
some regions of the state space are not available, e.g.\ because the firm's
budget does not suffice to buy the respective labour.

\begin{definition}[Viable region]
At time $t$, the set of all economic states that could be reached at time $t+1$
without violating the budget limitations of an agent is called the viable region
at time $t$. An element of the viable region is called viable at time
$t$.\footnote{An association with viability theory (e.g. \cite{aubin1991}) is
intended although the detailed discussion of how it is connected with this work
is not part of the paper.}
\end{definition}

\begin{figure}
\begin{center}
\includegraphics[width=11cm]{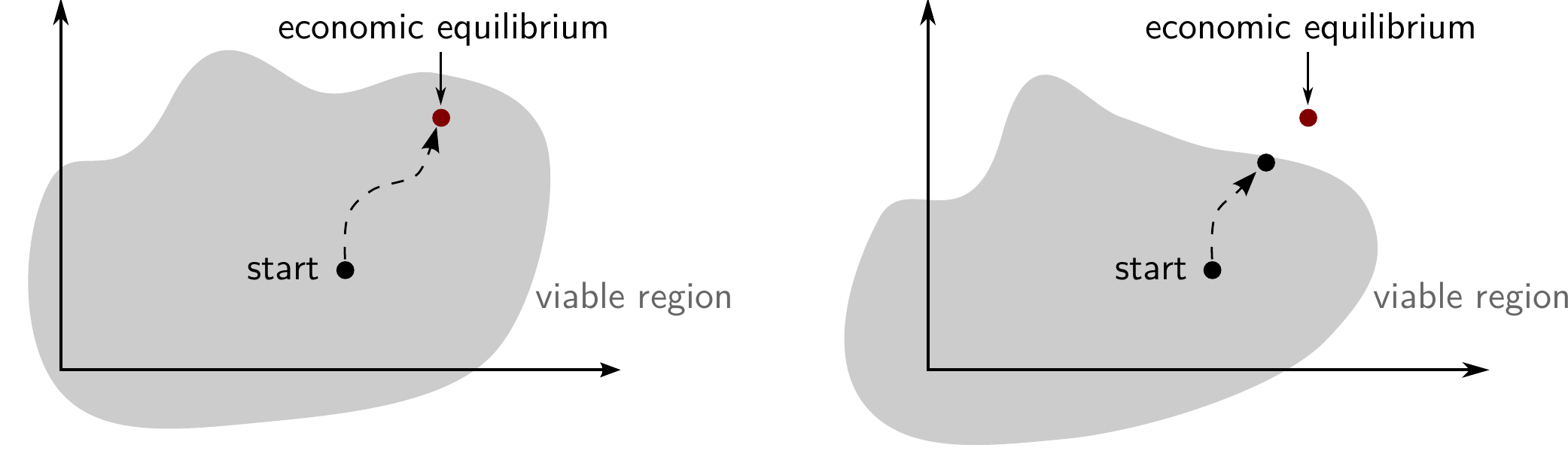}
\caption{Economic space with budget limitations: If the \textit{learning
trajectory} (broken line) towards the economic equilibrium lies completely
inside the viable region, the economic equilibrium is obtained eventually (left). If not, a
dynamic equilibrium (a \textit{border equilibrium}) in reached at the border of
the viable region (right).
\label{F:stateSpace}}
\end{center}
\end{figure}

The statement of Proposition \ref{P:types} is sketched in Figure
\ref{F:stateSpace}. There can be two possible situations: 1. The learning
trajectory lies completely in the viable region at all times $t$. Then, as
Proposition \ref{P:convergence} states, the economic equilibrium is obtained eventually.
2. The learning trajectory does not lie completely in the viable region.
Then the system evolves until it reaches the boarder of the viable region and
stays there.\\

The viable region can vary over time depending on how the budget limitations of
the agents develop over time.

\begin{proposition}\label{P:shrinkingViability}
If there is no money, economic states that are not viable in the beginning
cannot become viable later, i.e. the viable region can only shrink over time.
\end{proposition}

Proof: If there is no money, the firm's budget $B^{(F)}_{t+1}$ is given by the
amount of jelly $J_t$ sold at time $t$ times the jelly price $p_t$, i.e.
$B^{(F)}_{t+1}= p_t J_t$. The household's budget is its current income,
i.e. $B^{(H)}_{t}=L_t$ (since $w_t=1$). That means the household's
budget limitation is $p_t J_t\le L_t$, and the firm's budegt limitation is
$L_{t+1}\le p_t J_t$. From these two inequalities follows $p_{t+1} J_{t+1} \le
p_{t} J_{t}$ and $L_{t+1}\le L_t$, i.e. $B^{(F)}_{t+1}\le B^{(F)}_{t}$ and
$B^{(H)}_{t+1}\le B^{(H)}_{t}$. But if the budgets of both agents cannot become
larger, neither does the viable region. $\qed$\\

\begin{definition}[Viability closure]
For a system with a constant total amount of money $M = m_t^{(H)}+m_t^{(F)}$,
the set of economic states that are viable when both agents have $M$ at their
disposal at the time they make decisions about supply, demand, and prices, is
called viability closure.
\end{definition}

\begin{proposition} \label{P:viabilityClosure}
In a system with money, the viability closure is constant over time and the
viable region is a subset of the viability closure.
\end{proposition}

Proof: The viability closure is constant over time since the total amount of
money in the system $M=m^{(H)}+m^{(F)}$ is constant over time. Since
the firm's budget $B^{(F)}_{t}=m^{(F)}_{t}\le M$ and the household's budget
$B^{(H)}_{t}=m^{(H)}_{t}\le M$ all viable states at time time lie in the
viability closure. $\qed$\\

With ``full" money, on the other hand, it is not so clear how the viable region
changes over time as money can be redistributed among agents. However, there is an upper
limit for the size of the viable region, given by the (constant) total amount of
money in the system.

\begin{proposition}\label{P:minMoney}
For a system with total amount of money $M$ and economic-equilibrium labour
$L_E$, if $M<L_E$ in the beginning, the economic equilibrium can not be reached.
\end{proposition}

Proof: To employ $L_E$ units of labour at time $t$, the firm needs to be able to
pay them, thus $m^{(F)}_t \ge L_E$ has to hold. But if $M<L_E$ this is not
possible because $m^{(F)}_t \le M$ and thus $m^{(F)}_t < L_E$. $\qed$\\

%=======================================================================================

\section{Simulations}
\label{S:simulations}
 
We programmed a model with the characteristics described above to illustrate the
dynamic evolution of such a model discussed in the previous section. As
simulation outcome, we are particularly interested in whether the system finally
obtains the economic equilibrium or not, which can be observed by either
comparing supply and demand on both markets or by recording the final utility.

\subsection{Final Utility Versus Initital Expectations With and Without Money}

\begin{figure}[p]
\begin{center}
\includegraphics[width=11.6cm]{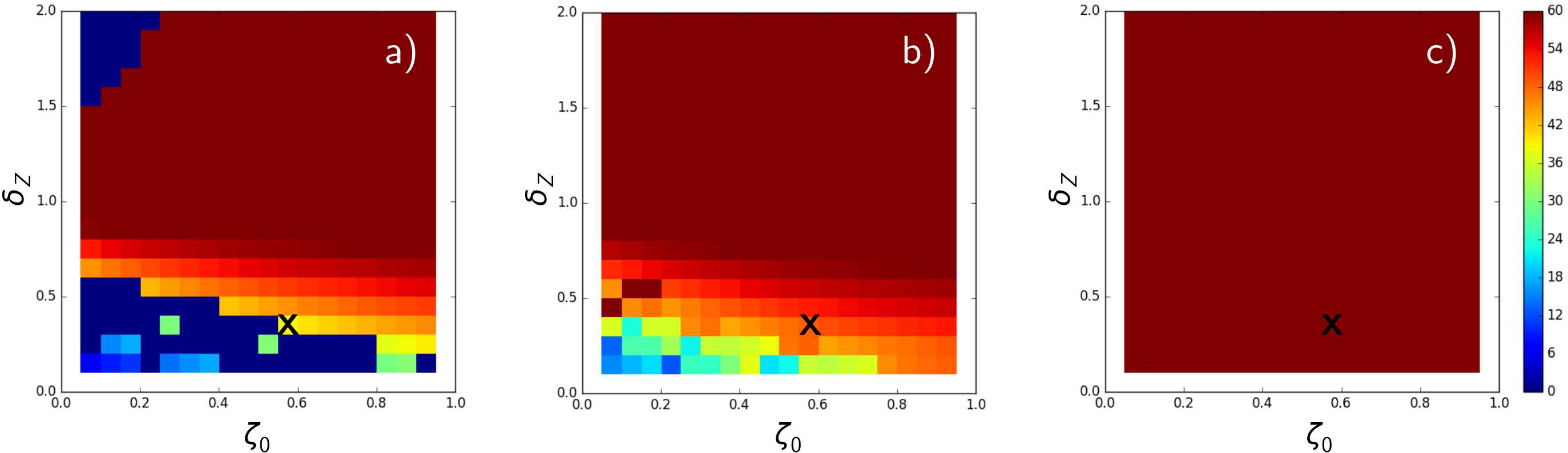} 
\caption{Utility after 50 time steps versus initial demand expectations of the
firm, $\zeta_0$ and $\Delta_z$. a) No money and ``right'' initial amount of
revenue by the firm. b) With money; initial amount of firm corresponds to
initial revenues in a), household with no initial money. c) Firm and household
have a large amount of money initially. Temporal evolution of the economic
variables at $\zeta_0=0.55$ and $\Delta_z=0.3$ (marked point) are shown Figures
\ref{F:trajectories-mon0}, \ref{F:trajectories-mon1-onlyF}, and
\ref{F:trajectories-mon1-both400}.
\label{F:utilityVsExp}}
\end{center}
\end{figure}

\begin{figure}[p]
\begin{center}
\begin{tabular}{c c c}
\includegraphics[width=4.14cm]{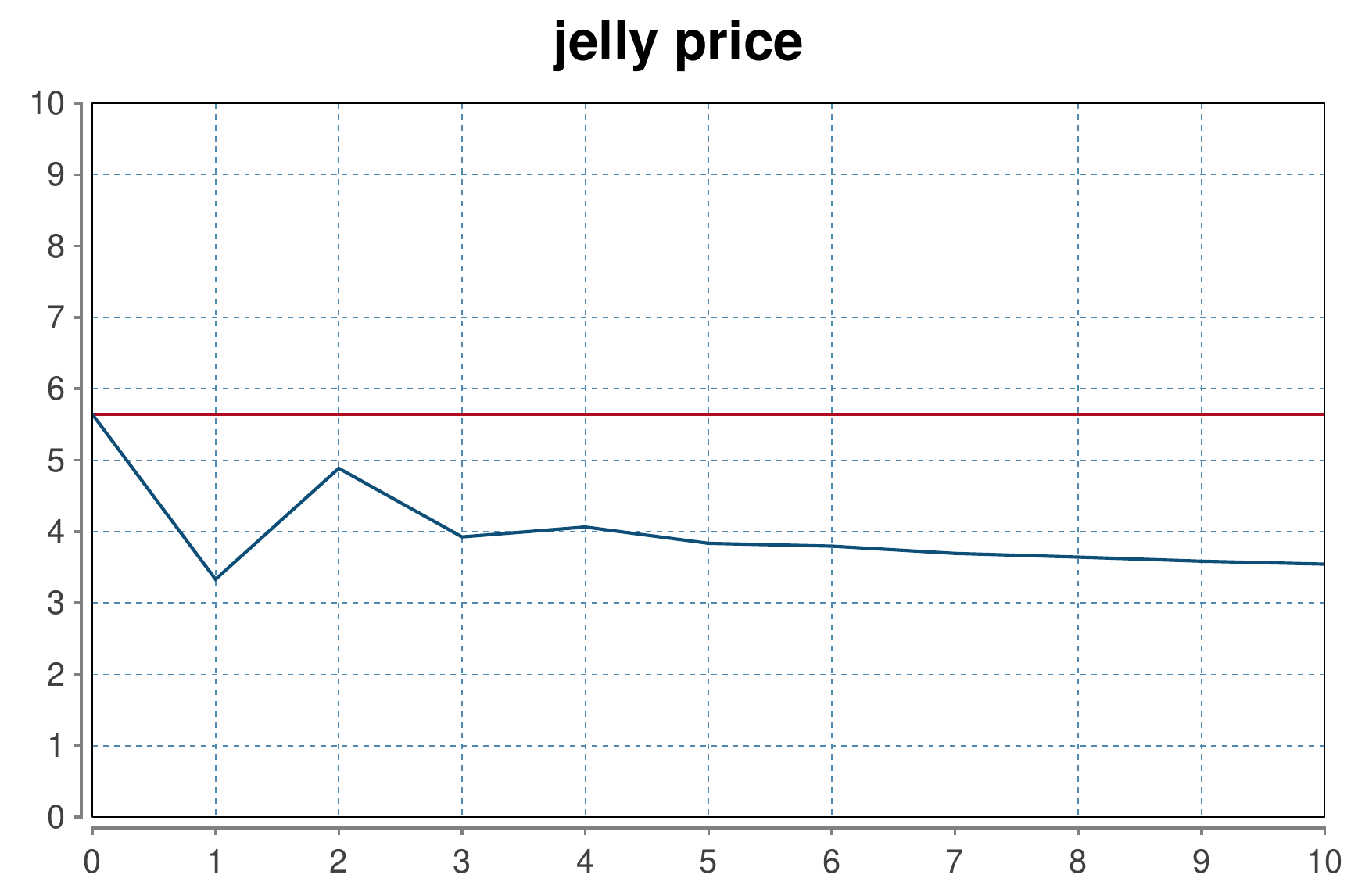} &
\includegraphics[width=4.14cm]{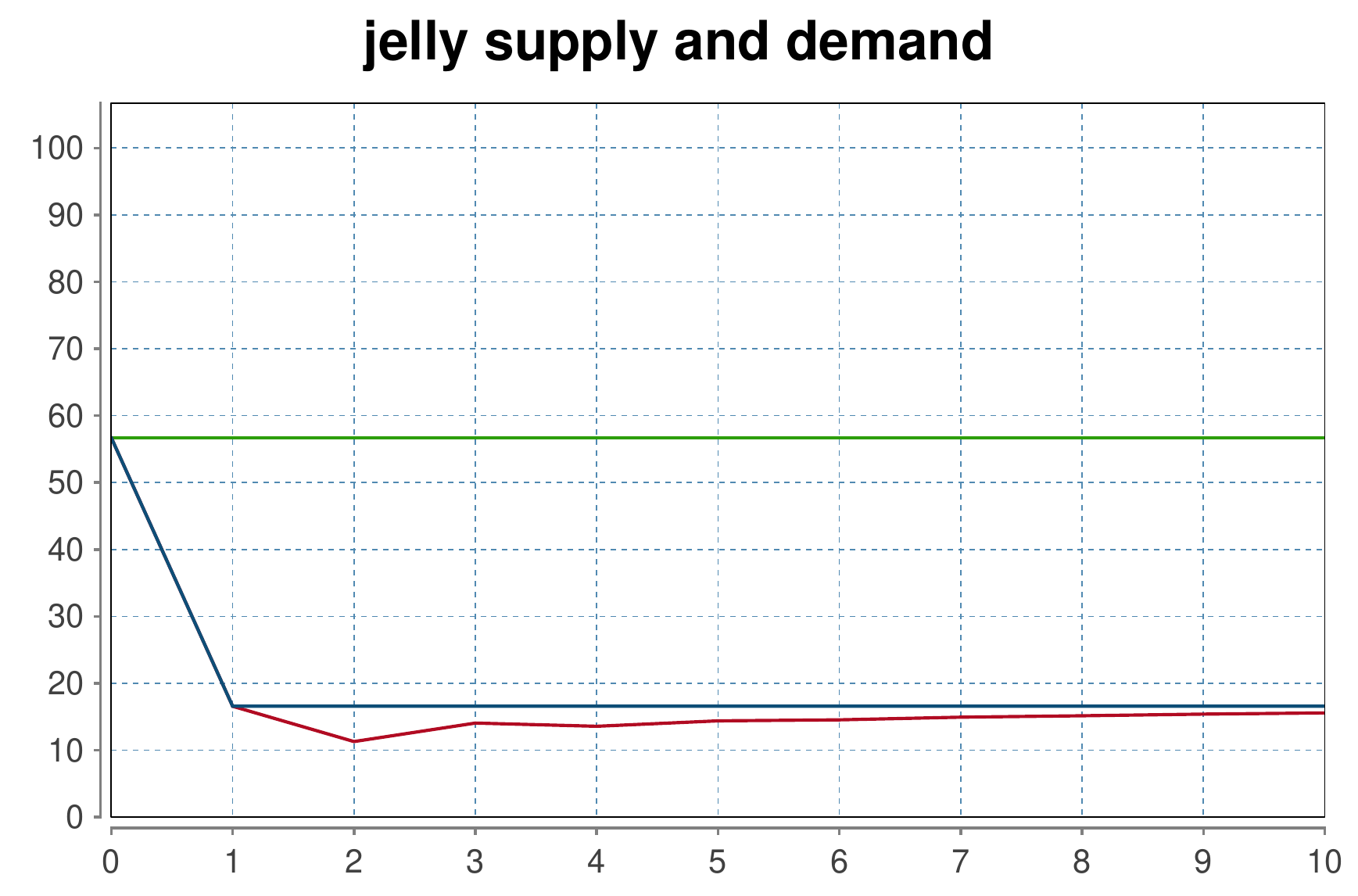} &
\includegraphics[width=4.14cm]{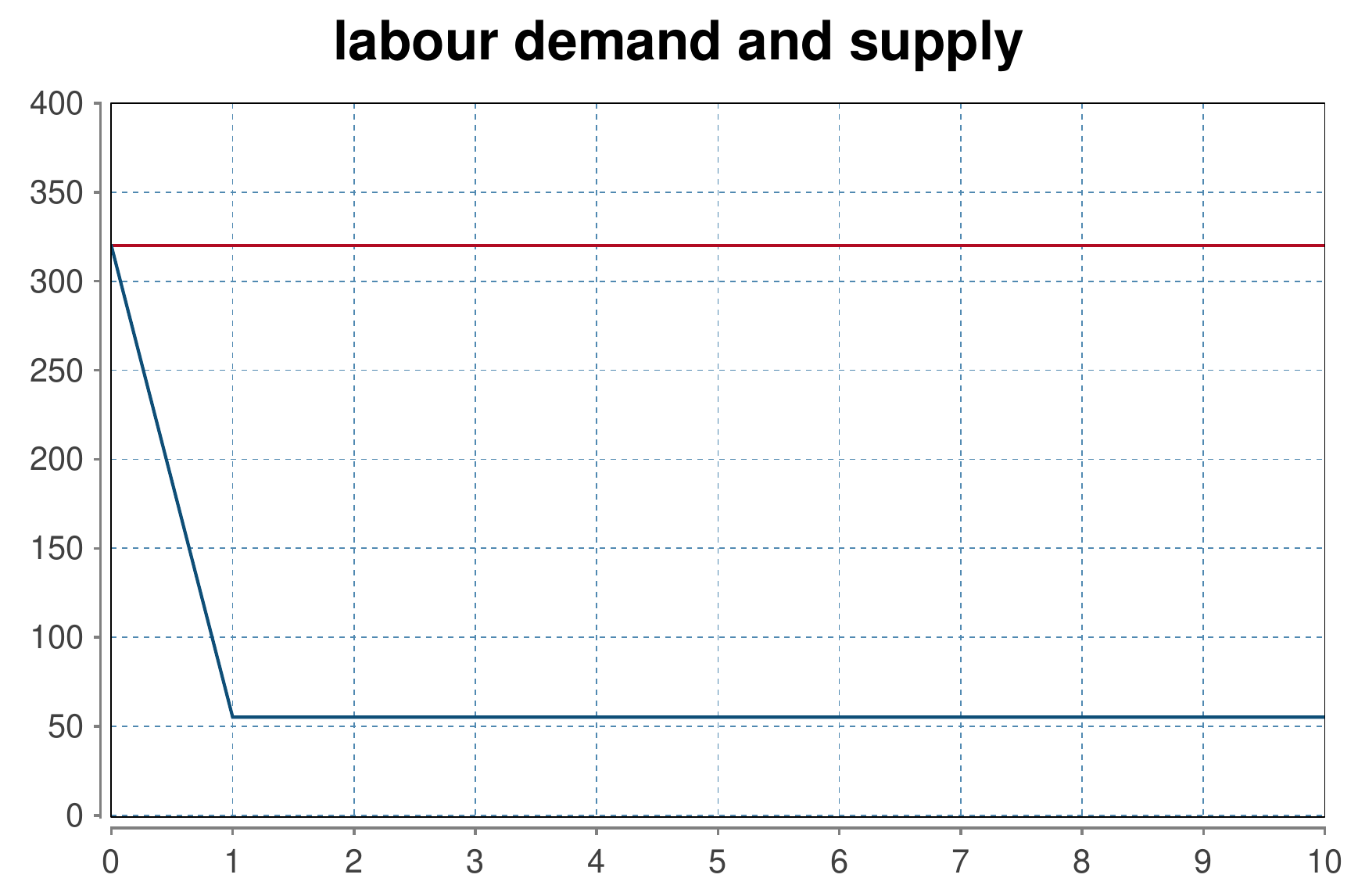}
\end{tabular}
\caption{Temporal evolution of system with initial expectations $\zeta_0=0.55$
and $\Delta_z=0.3$ from Figure~\ref{F:utilityVsExp}a (border equilibrium).
From left to right: a) Jelly price (blue) and price at economic equilibrium
(red). b) Jelly supply (blue), demand (red), and demand and supply at economic
equilibrium (green). c) Labour demand (blue) and supply (red); the labour supply
also equals demand and supply at economic equilibrium.
\label{F:trajectories-mon0}}
\end{center}
\end{figure}

\begin{figure}[p]
\begin{center}
\begin{tabular}{c c c}
\includegraphics[width=4.14cm]{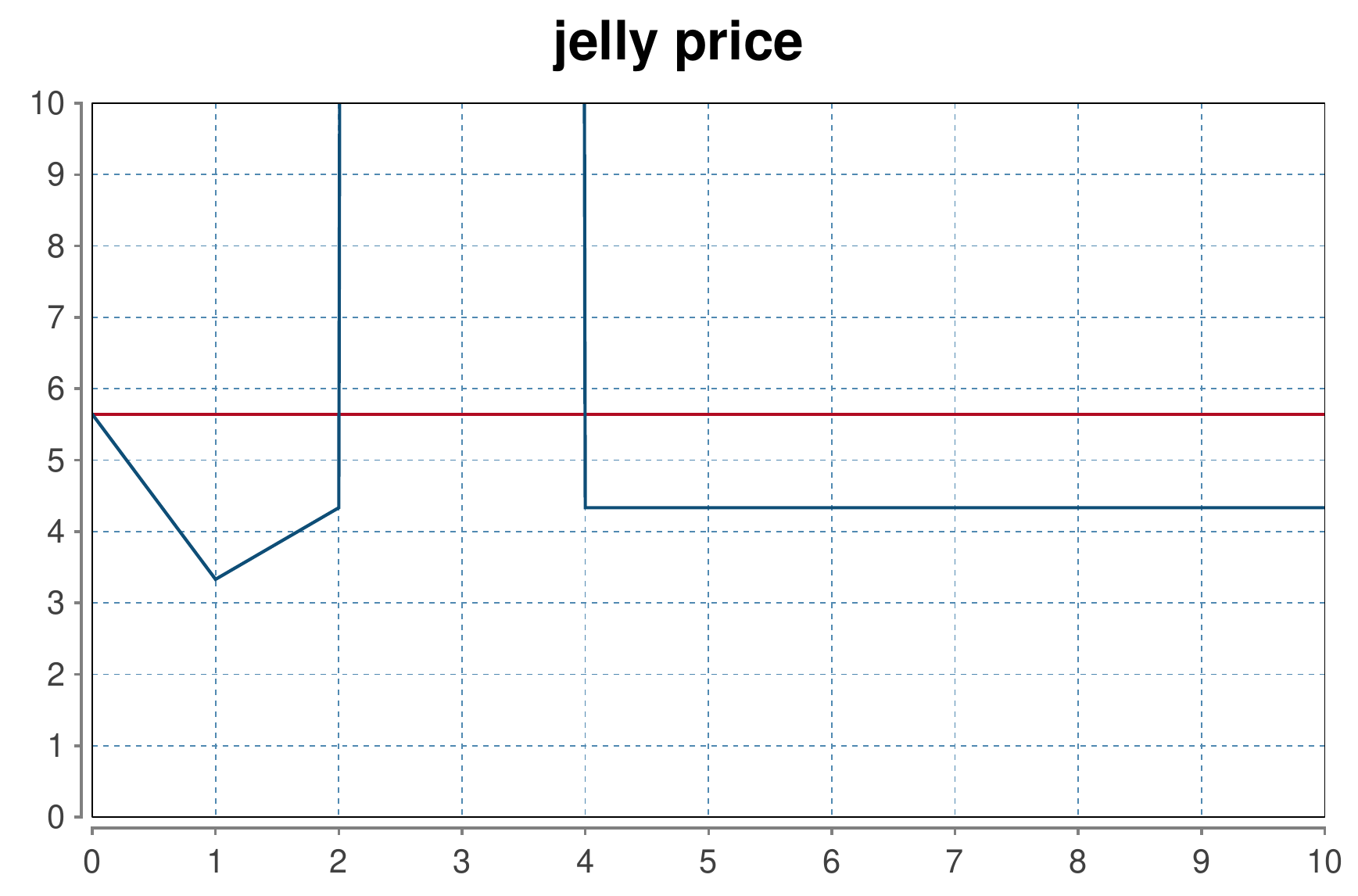} &
\includegraphics[width=4.14cm]{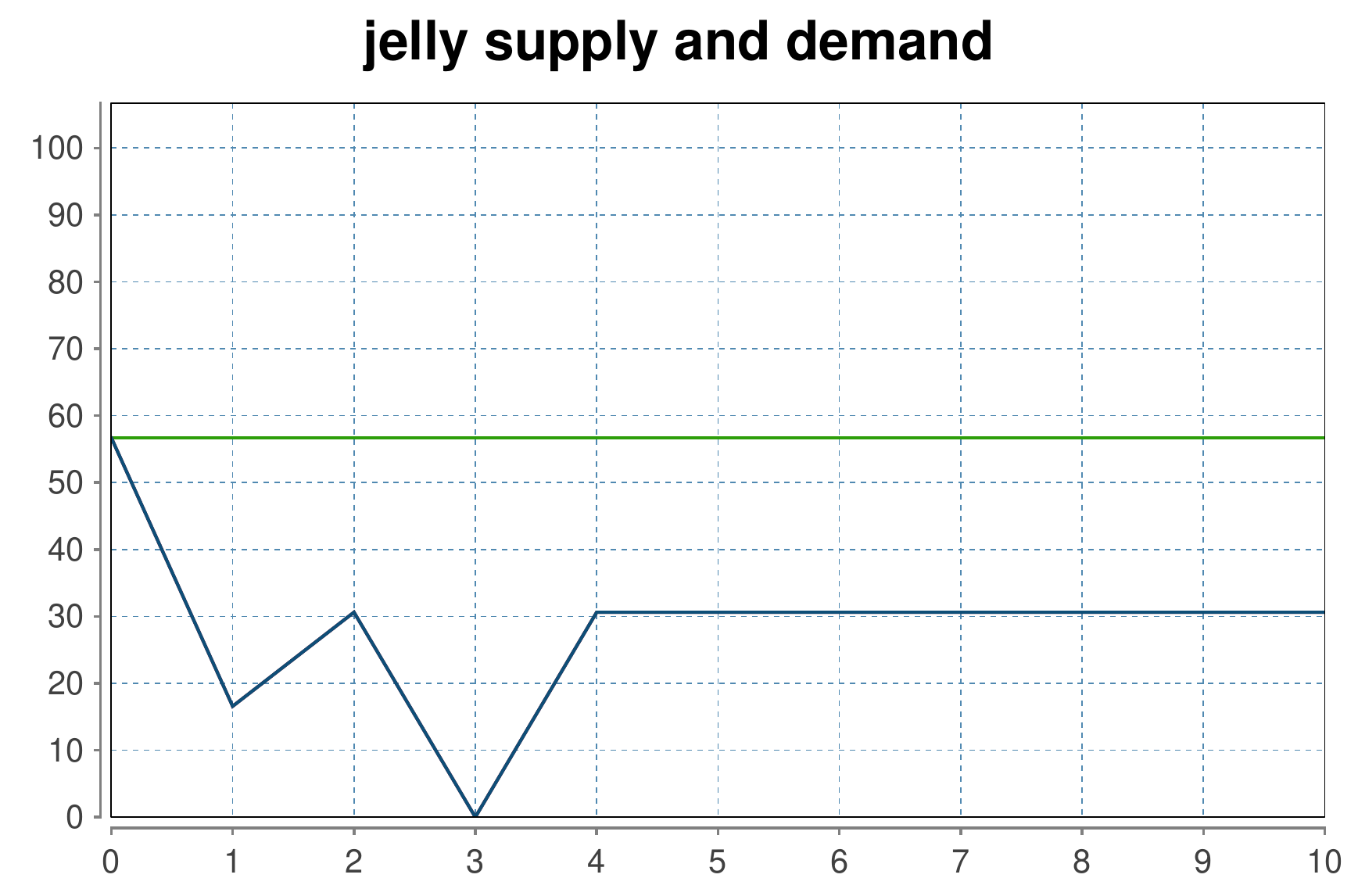} &
\includegraphics[width=4.14cm]{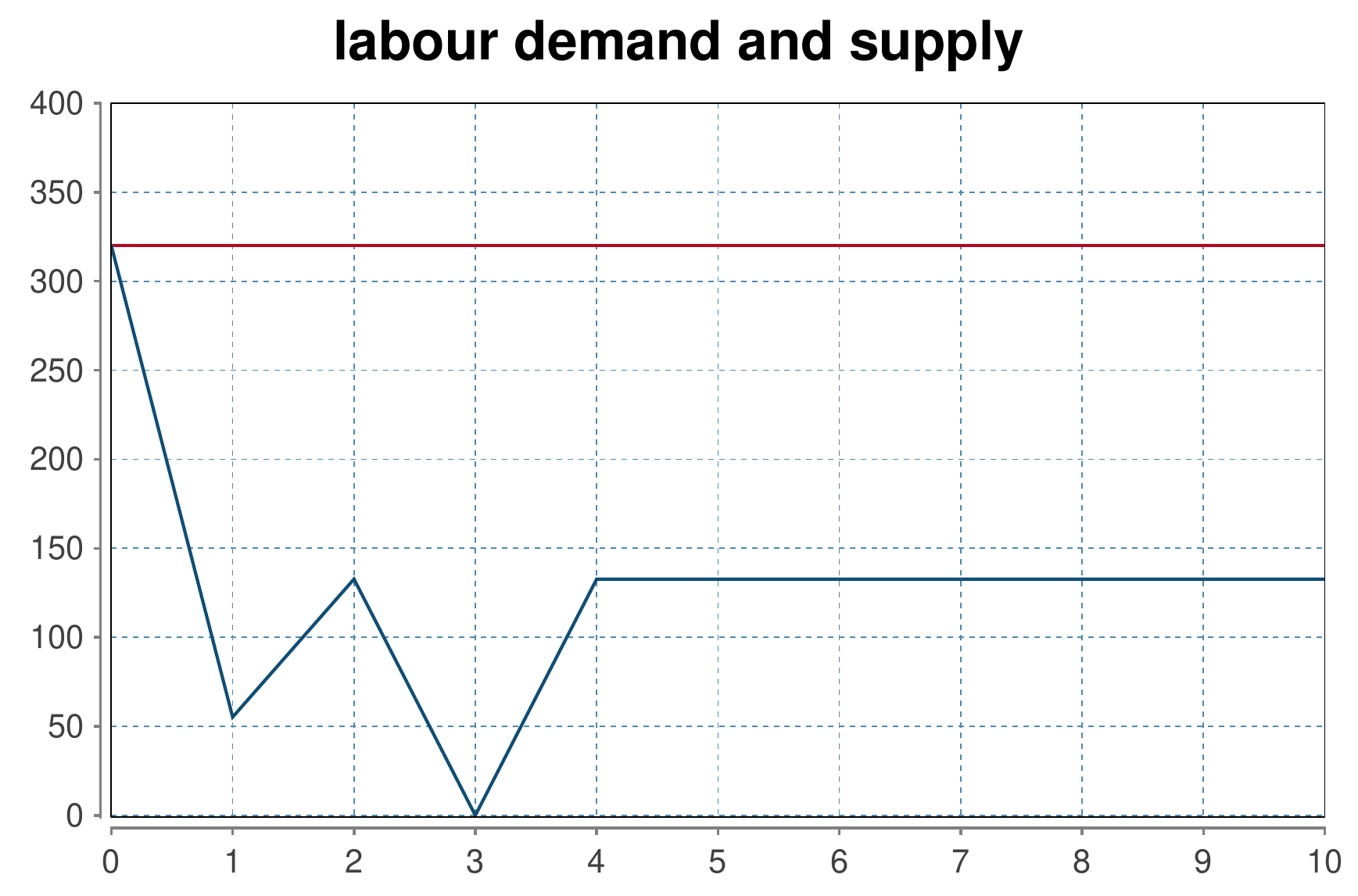}
\end{tabular}
\caption{Temporal evolution of system with initial expectations $\zeta_0=0.55$
and $\Delta_z=0.3$ from Figure~\ref{F:utilityVsExp}b (border equilibrium).
From left to right: a) Jelly price (blue) and price at economic equilibrium
(red). b) Jelly supply (blue), demand (red, under blue line), and demand and
supply at economic equilibrium (green). c) Labour demand (blue) and supply
(red); the labour supply also equals demand and supply at economic equilibrium.
\label{F:trajectories-mon1-onlyF}}
\end{center}
\end{figure}

\begin{figure}[p]
\begin{center}
\begin{tabular}{c c c}
\includegraphics[width=4.14cm]{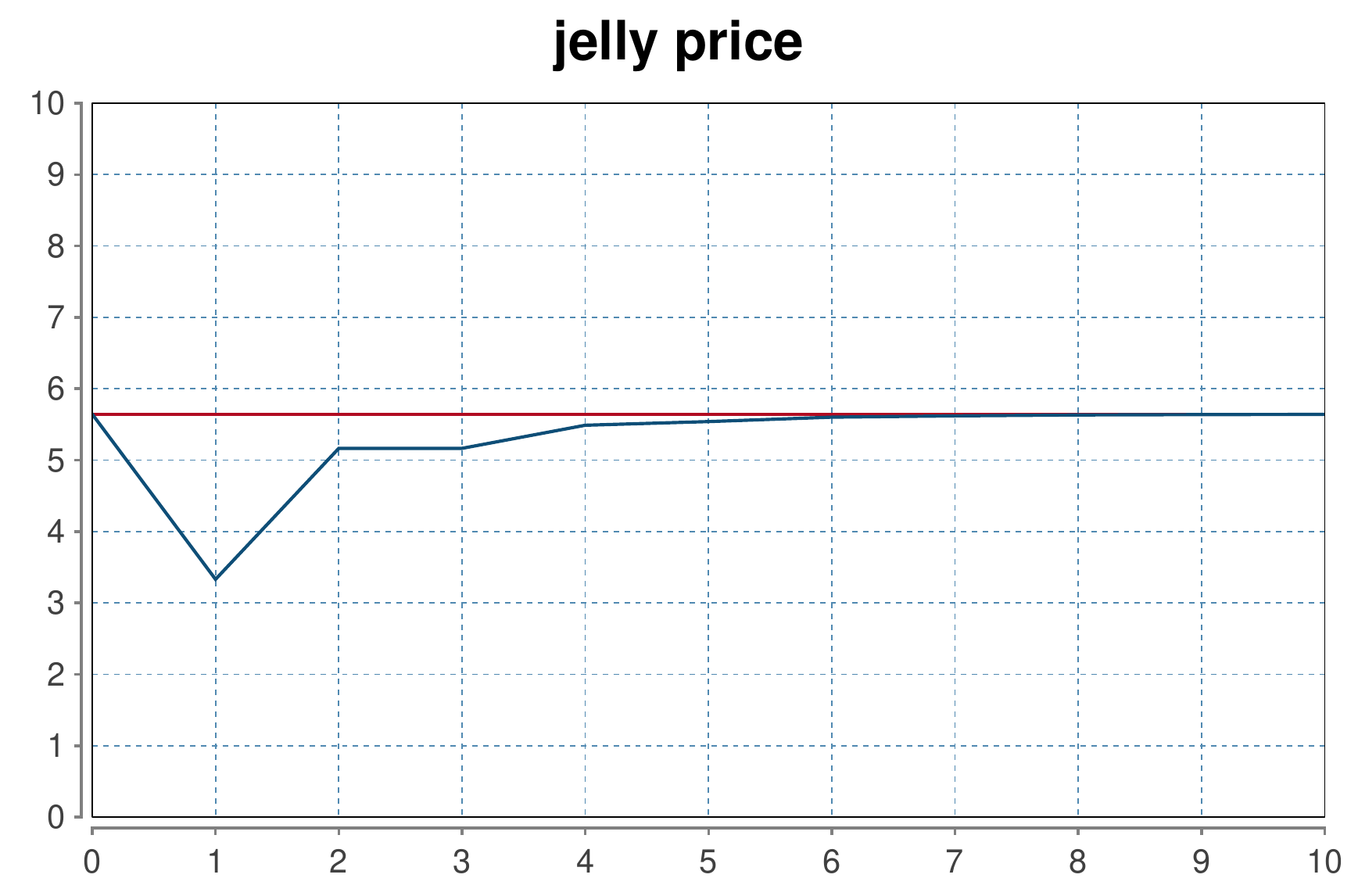} &
\includegraphics[width=4.14cm]{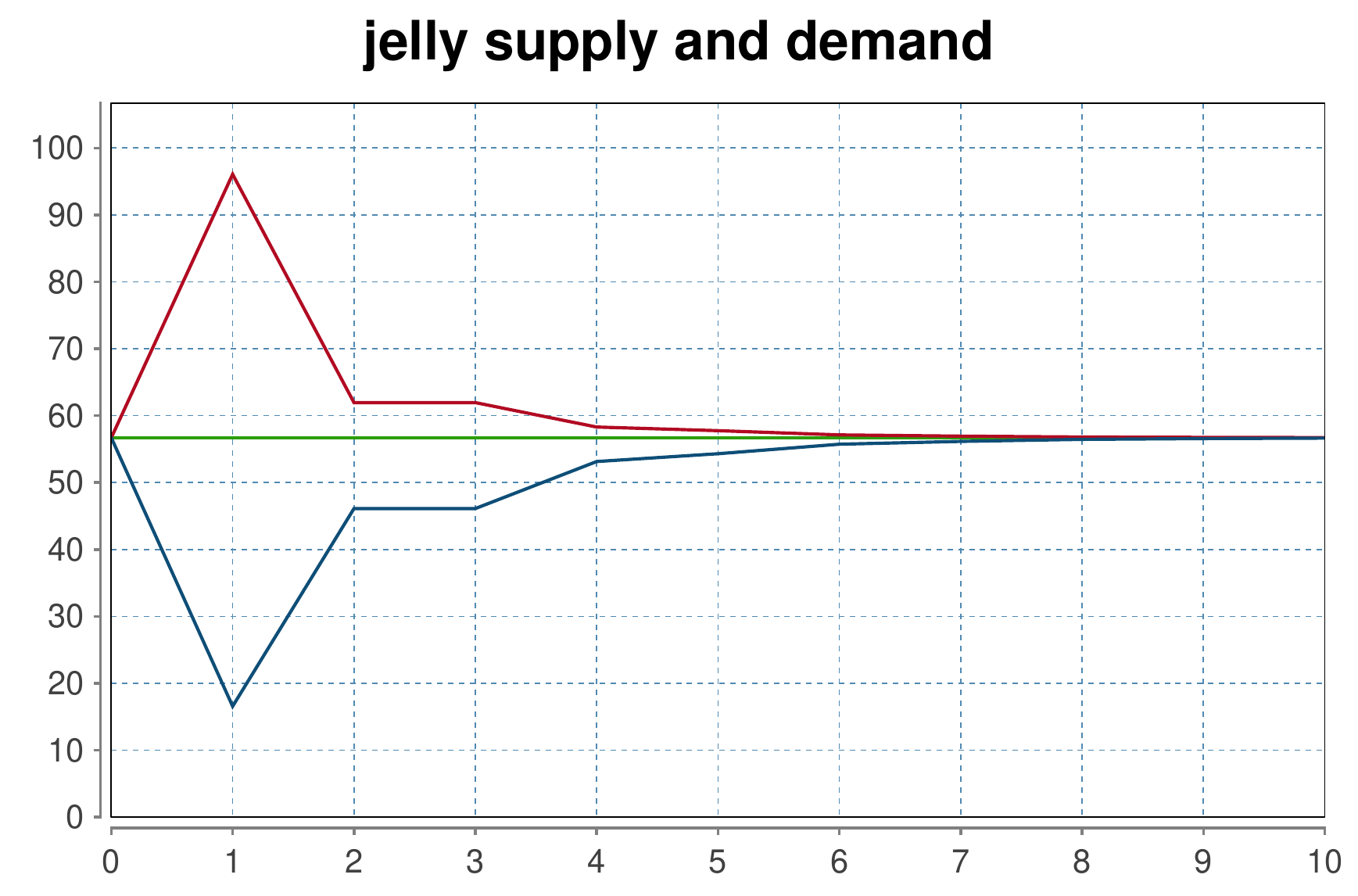} &
\includegraphics[width=4.14cm]{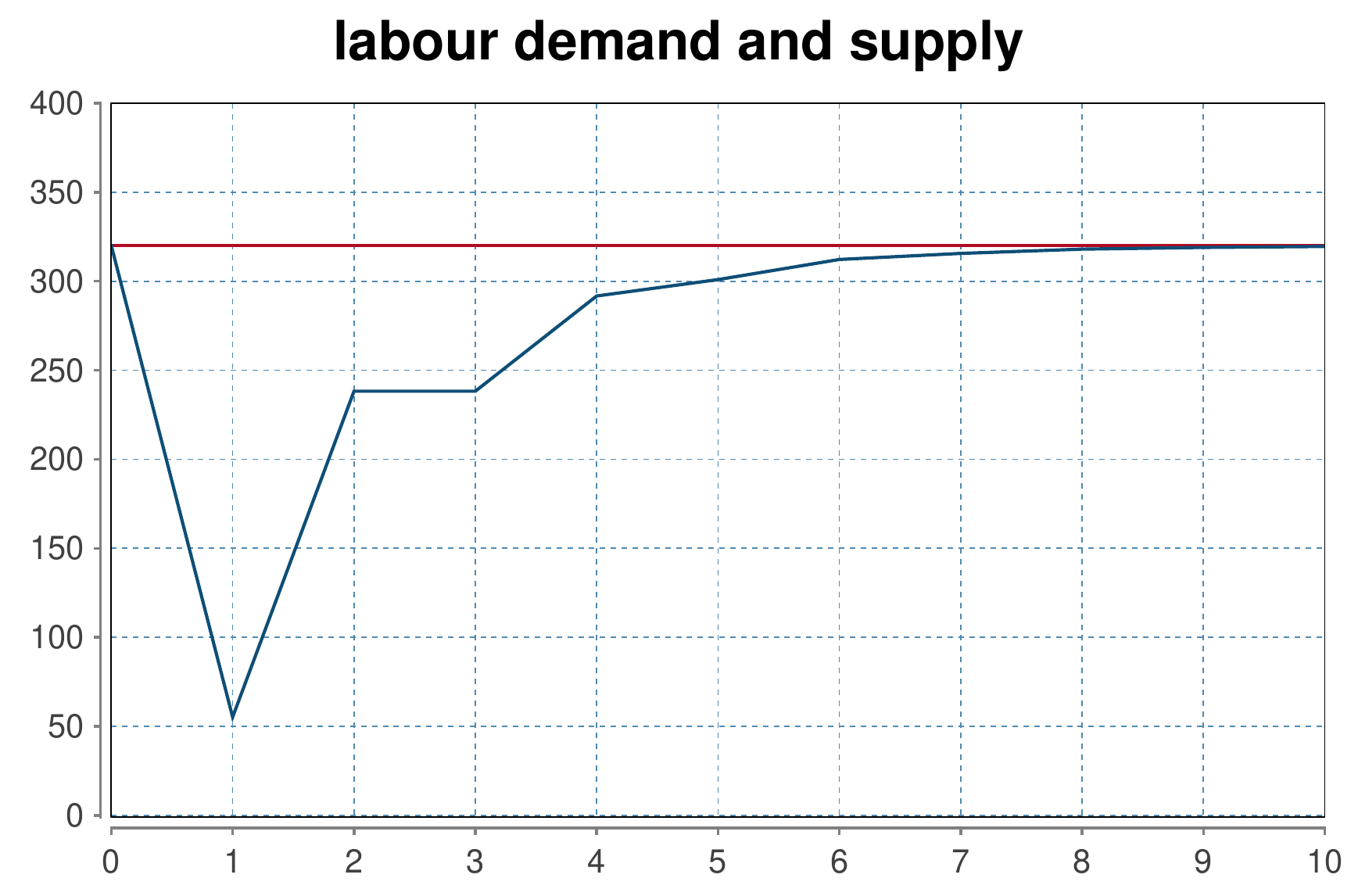}
\end{tabular}
\caption{Temporal evolution of system with initial expectations $\zeta_0=0.55$
and $\Delta_z=0.3$ from Figure~\ref{F:utilityVsExp}c (economic equilibrium).
From left to right: a) Jelly price (blue) and price at economic equilibrium
(red). b) Jelly supply (blue), demand (red, under blue line), and demand and
supply at economic equilibrium (green). c) Labour demand (blue) and supply
(red); the labour supply also equals demand and supply at economic equilibrium.
\label{F:trajectories-mon1-both400}}
\end{center}
\end{figure}

Figure \ref{F:utilityVsExp} shows the utility at the fixed point the system
converges to (utility after 50 time steps) versus the initial parameter of the
firm's expected demand function. In Figure \ref{F:utilityVsExp}a) simulations
without money are shown, where the initial revenue (at $t=0$) of the firm is
supposed to be exactly what it would be at the economic equilibrium (i.e. the
economic equilibrium is viable at $t=1$). The firm's initial expectations are
characterised by $\zeta_0$ and $\Delta_z$. The firm's expected demand function
is given by
\begin{equation}
\phi_t (J) = \frac{z_t}{J^{\zeta_t}}.
\end{equation}
$\zeta_0$ is the initial value for $\zeta$. Let $p_E$ and $J_E$ be jelly price
and jelly demand at the economic equilibrium. $\delta_z$ characterises how much
the initial $z_0$ deviates from the value $\tilde{z}=p_E\cdot J_E^{\zeta_0}$,
i.e. from the value for which, given $\zeta_0$, $\phi_0(J_E)=p_E$. So the
initial $z_0$ is given by $z_0=p_E J_E^{\zeta_0}\cdot \delta_z$.
 
Figure \ref{F:utilityVsExp}b) also shows the utility after 50 time steps versus
initial expectations $\zeta_0$ and $\Delta_z$, but this time with money: the
initial cash balance of the firm equals the firm's revenues at economic
equilibrium, i.e. in the first time step the firm can buy as much labour as in
Figure \ref{F:utilityVsExp}a), and the economic equilibrium is as well viable at
time $t=1$. The initial cash balance of the household is zero, so the initial
situation is quite similar to the one of Figure \ref{F:utilityVsExp}a), only
that there is money which can store value over time. The final utility looks
similar to Figure \ref{F:utilityVsExp}a), however, for some $\zeta_0$ and
$\Delta_z$ combinations the final utility is higher and the system reaches the
economic equilibrium more often.

For the simulations of Figure \ref{F:utilityVsExp}c) firm and household have a
have large amount of money at their disposal initially:
$m^{(F)}_0=m^{(H)}_0=400$ (which is equivalent to the labour force,
since $L_f=400$ in all the simulations). So there are basically no budget
limitations, and as stated by Proposition \ref{P:convergence} the system
converges to the economic equilibrium for all initial expectations $\zeta_0$,
$\Delta_z$.

Figures \ref{F:trajectories-mon0}, \ref{F:trajectories-mon1-onlyF}, and
\ref{F:trajectories-mon1-both400} show the temporal evolution of prices, and 
jelly and labour market for $\zeta_0=0.55$ and $\Delta_z=0.3$, i.e. the point
marked in Figures \ref{F:utilityVsExp}a-c.

\subsection{Initial Total Money and Initial Money Distribution}

\begin{figure}
\begin{center}
\includegraphics[width=5.1cm]{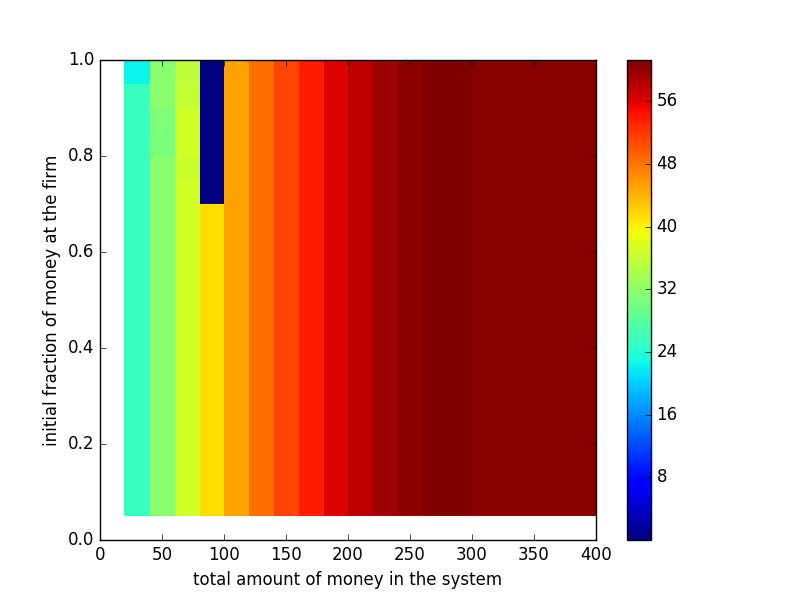}
\caption{Utility after 50 time steps versus amount of total money in the system
$M$ and initial money distribution. The initial money distribution hardly
matters, but the total amount of money is important. For $M\ge L_E =320 $, the
economic equilibrium is obtained.
\label{F:utilityVsMon}}
\end{center}
\end{figure}

\begin{figure}
\begin{center}
\begin{tabular}{c c c}
\includegraphics[width=4.14cm]{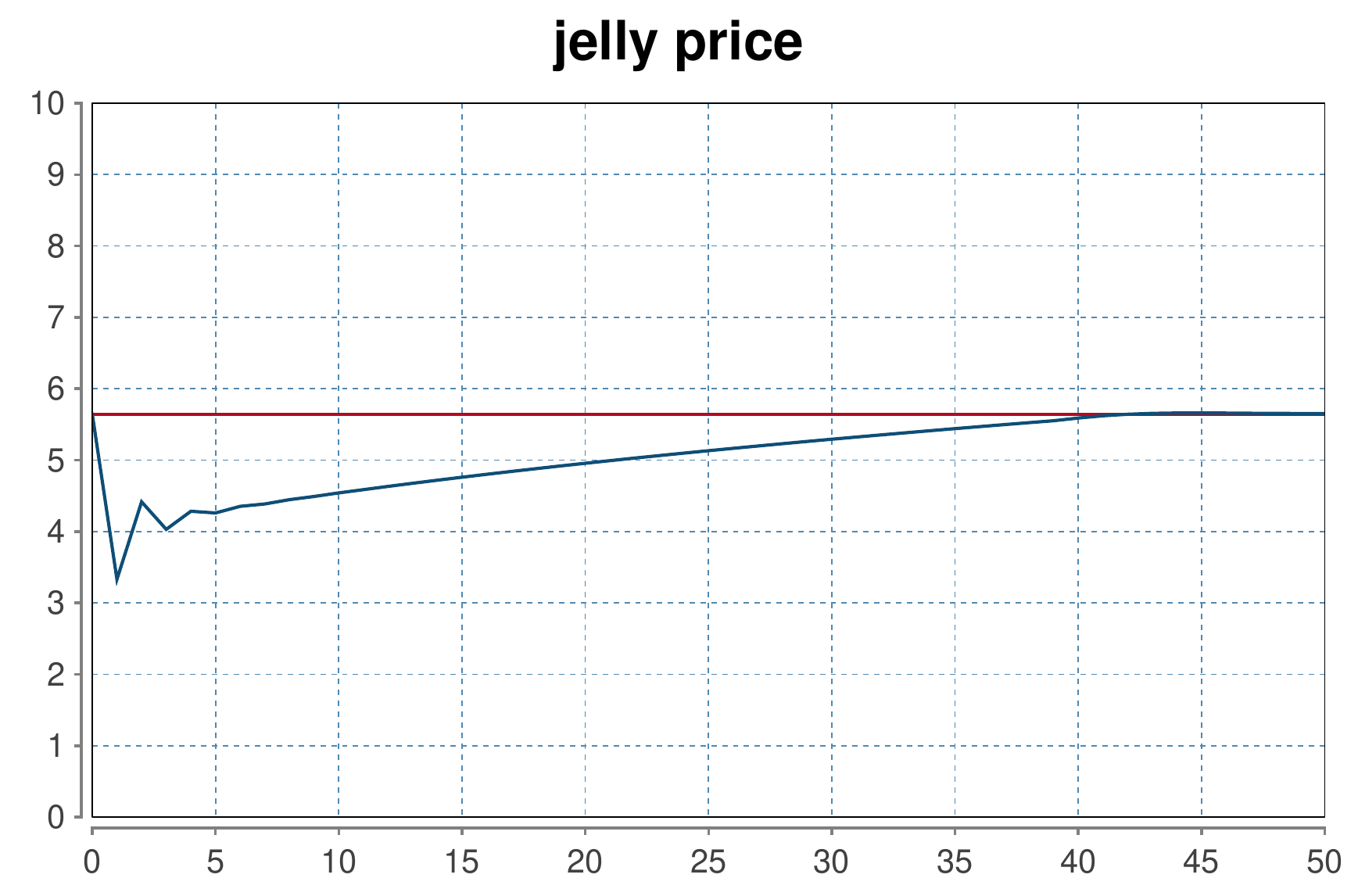} &
\includegraphics[width=4.14cm]{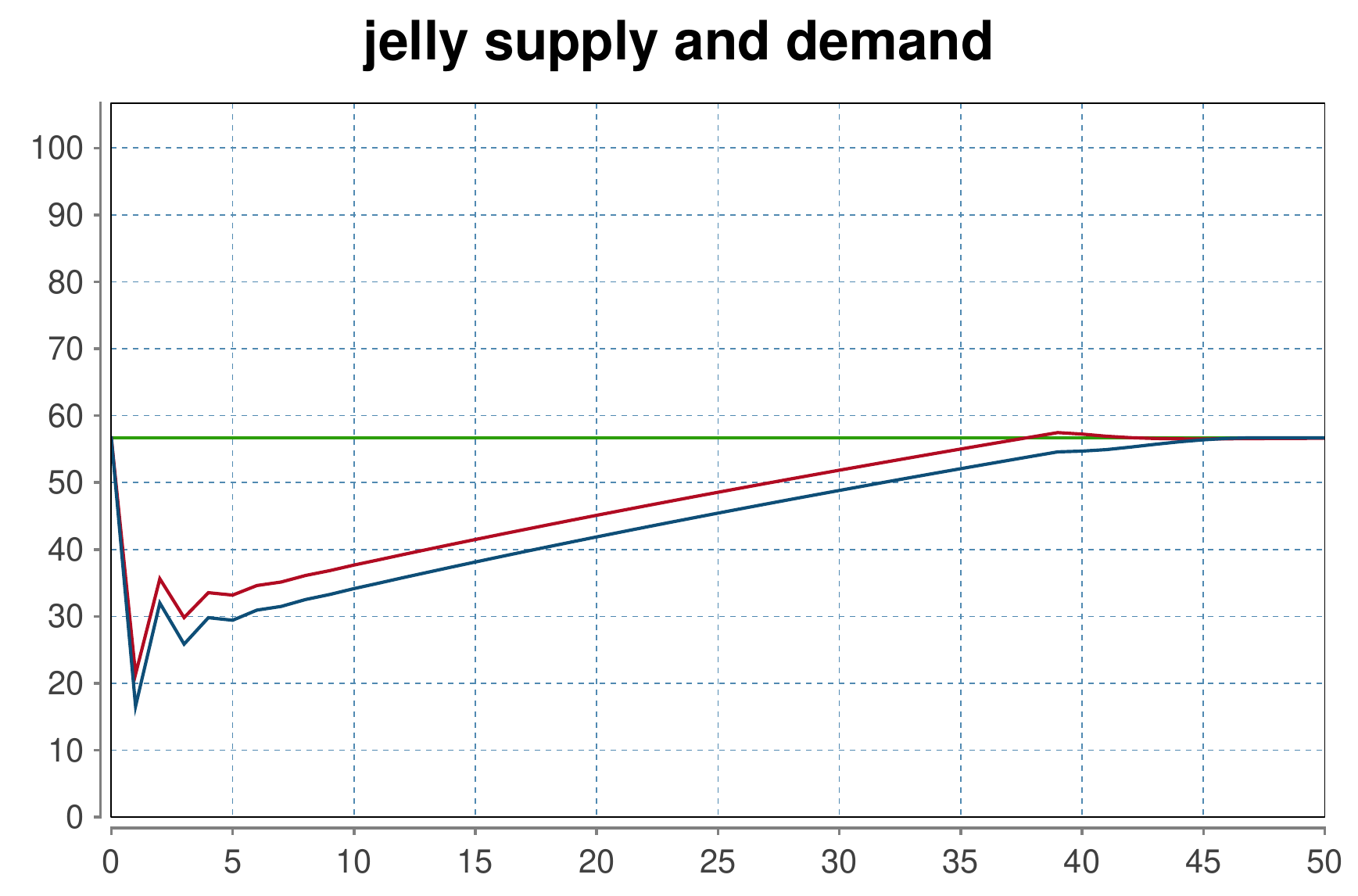} &
\includegraphics[width=4.14cm]{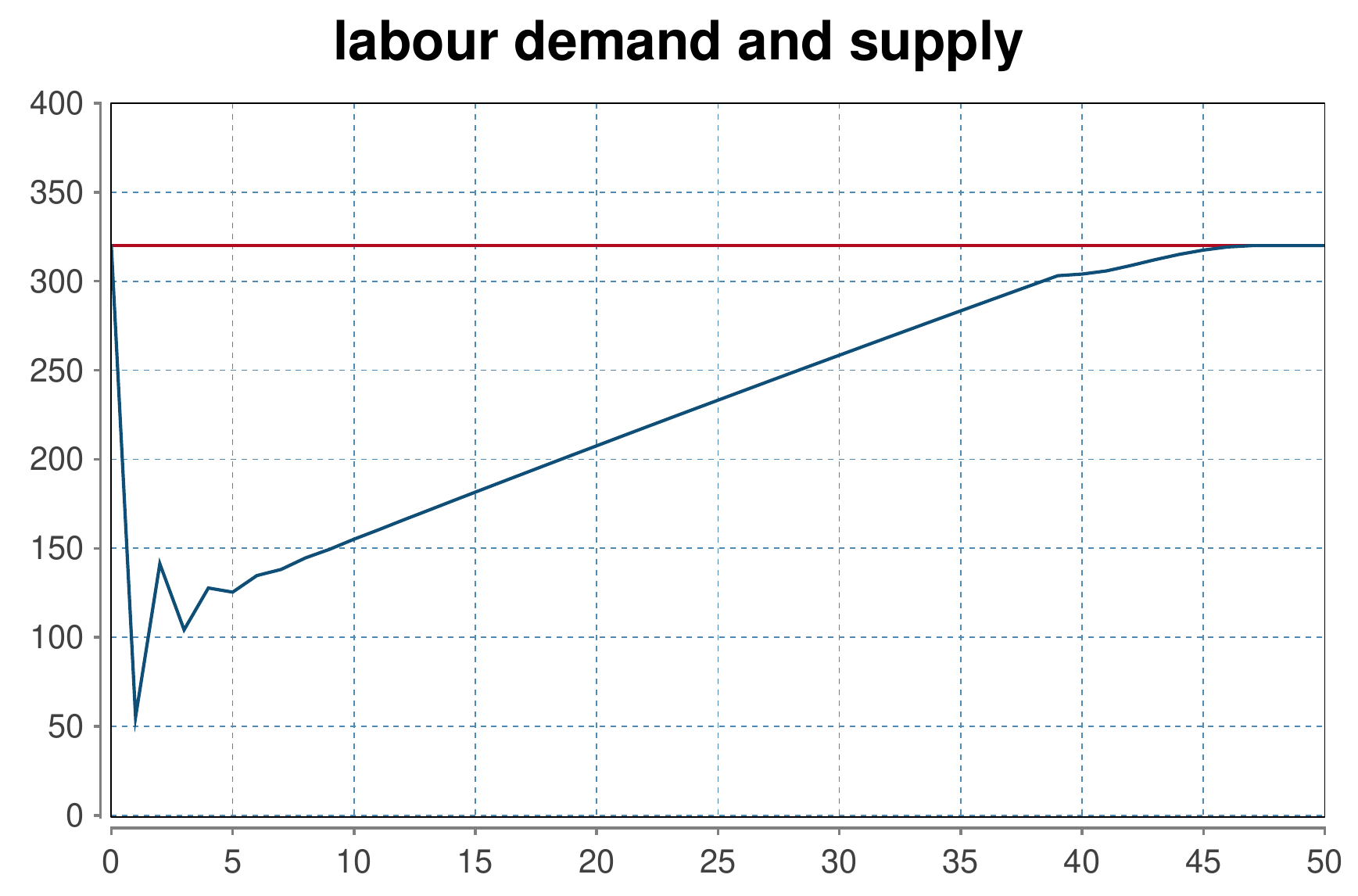}
\end{tabular}
\caption{Temporal evaluation as in Figure \ref{F:trajectories-mon1-onlyF} with
the only difference that the firm does not posses all the money initially but
about 5\% is initially with the household.
From left to right: a) Jelly price (blue) and price at economic equilibrium
(red). b) Jelly supply (blue), demand (red), and demand and supply at economic
equilibrium (green). c) Labour demand (blue) and supply (red); the labour supply
also equals demand and supply at economic equilibrium. 
\label{F:trajectories-mon1-95F}}
\end{center}
\end{figure}

For the case with money, Figure \ref{F:utilityVsMon} shows the final utility
versus the total amount of money $M$ in the system and the initial money
distribution. In the simulation $L_E=320$. It can be seen that, as stated by
Proposition \ref{P:minMoney}, if $M < L_E$ the economic equilibrium is not
obtained. In contrast to the total amount of money $M$ in the system, the
initial money distribution seems to be quite irrelevant for the long term
equilibrium. But in some cases it can make a difference: Figure
\ref{F:trajectories-mon1-95F} shows the same simulation as Figure
\ref{F:trajectories-mon1-onlyF} with the only difference that the firm does not
posses all the money initially but around 5\% of it is with the household.
Unlike in the simulation from Figure \ref{F:trajectories-mon1-onlyF}, here the
economic equilibrium is obtained.

%=======================================================================================

\section{Model Extension: Money As a Third Commodity}
\label{S:realMoney}

The three versions of our model (see Section \ref{S:money}) differ in whether
and how money is part of the economic system. Up to now, only the dynamics of
the first two of them has been discussed (Sections \ref{S:dynamics} and
\ref{S:simulations}). While in the first version there is no money at all, in
the second version there is -- but neither household nor firm have a demand for
it. It does not play a role when agents do their optimisations to plan supply
and demand of labour and jelly. Only when it comes to whether their plans can be
fulfilled their cash balances may be important.

In this Section, the third version is discussed: it is assumed that the
household wants to keep a certain amount of its income in cash. That means that a demand
for money is created. But as soon as there is a demand for money, money has a
price and can serve as \num.

The evolution rules for the extended system are the same as before (see Section
\ref{S:evoRules}). Only here, when the firm sets the jelly price it also sets
the wage, and when updating its expected jelly demand function it also updates its
expected labour supply.

\subsection{Agents' Optimisations}

For determining supply and demand, the household optimises its utility and the
firm its profit. Here, there are two changes compared to the previous versions
(Sections \ref{S:household} and \ref{S:profitOpt}), firstly, the household's
budget constraint takes into account the preference of having a certain amount
of money in cash (as described in Section \ref{S:household2}) and secondly,
since money now serves as \num, the firm has to set a wage which is done based
on its labour supply expectations (Section \ref{S:profitOpt2}) .

\subsubsection{Household}
\label{S:household2}

The household is assumed (to plan) to keep a fraction $s$ of its current income
in cash (e.g. to cover unforeseen costs). The budget constraint becomes
\begin{eqnarray}
p J +s \cdot wL &=& w L + m^{(H)}, \;\;\;\; \text{that~is} \nonumber \\
p J  &=& (1-s) \cdot w L + m^{(H)}. \label{E:budgetConstraintM}
\end{eqnarray}
with $m^{(H)}$ being the monetary holdings of the household from the previous
time period. 
Maximising the utility given in (\ref{E:utility}) subject to this budget
constraint (\ref{E:budgetConstraintM}) yields a utility maximum at
\begin{eqnarray}
L_H &=& \frac{\beta}{\alpha+\beta} \cdot L_f - \frac{\alpha}{\alpha+\beta}
\frac{m^{(H)}}{(1-s)\cdot w} \label{E:plannedLSM}\\
J_H &=& \frac{\beta}{\alpha+\beta} \cdot \left(  (1-s) \frac{w}{p} L_f +
\frac{m^{(H)}}{p} \right)\label{E:plannedJD}
\end{eqnarray}

\subsubsection{Firm}
\label{S:profitOpt2}

The firm's profit optimisation (see Section \ref{S:profitOpt}) has to be adapted
as well: With money being the \nume and assuming $\psi$ to represent the firm's
expectations about the labour supply (at wage $\psi(L)$ the firm expects the
labour supply to be $L$) it is given by
\begin{eqnarray}
	\min_{L}& &\vert pJ - wL \vert\\
	\text{s.t.}& &J = \rho(L) =  L^\gamma \\
	& & p = \phi(J) \\
	& & w = \psi(L)\label{E:wage}\\
	& & 0 \leq L \leq L_f,
\end{eqnarray}
that is
\begin{equation}
	\min_{0 \leq L \leq L_f} \vert \rho(L) \cdot \phi(\rho(L)) - L \cdot \psi(L)
	\vert. \label{E:profitOptL2}
\end{equation}

\subsection{Economic Equilibria}

If money is seen as a third commodity, economic equilibrium would mean
that not only in the labour and jelly market, but also in the money market
supply equals demand. However, in this section we only examine economic
equilibria in the same sense as for the previous model versions, i.e. we
suppose the economy to be in equilibrium if $L^{(S)}_t=L^{(D)}_t$ and
$J^{(S)}_t=J^{(D)}_t$, and then we are interested in whether these points are
fixed points in the economic space (as defined above) and whether there are
other fixed points. 

Economic equilibria can then be derived as follows. The household's utility
maximisation subject to its budget constraint yields a maximum at
\begin{eqnarray}
	L_H &=& \frac{\beta}{\alpha+\beta} \cdot L_f - \frac{\alpha}{\alpha+\beta}
	\frac{m^{(H)}}{(1-s)\cdot w}\\
	J_H &=& \frac{\beta}{\alpha+\beta} \cdot \left(  (1-s) \frac{w}{p} L_f +
	\frac{m^{(H)}}{p} \right)
\end{eqnarray} 
(see Section \ref{S:household2}). Necessary condition for an economic
equilibrium is $J_H = L_H^\gamma$, i.e. 
\begin{equation}
   \frac{\beta}{\alpha+\beta} \cdot \left(  (1-s) \frac{w}{p} L_f +
	\frac{M}{p} \right) = \left( \frac{\beta}{\alpha+\beta} \cdot L_f -
	\frac{\alpha}{\alpha+\beta}
	\frac{m^{(H)}}{(1-s)\cdot w} \right)^\gamma.
\end{equation}
For a dynamically stable equilibrium, however, also $m^{(H)}=swL$ has to hold
(otherwise the household would change its labour supply and jelly demand for
the next time step), that means for the equilibrium labour supply $L_E$
\begin{eqnarray}
	L_E &=& \frac{\beta}{\alpha+\beta} \cdot L_f - \frac{\alpha}{\alpha+\beta}
	\cdot \frac{s}{(1-s)} \cdot L_E, \;\; \text{and~thus} \nonumber \\
	L_E &=& \frac{\beta (1-s)}{\alpha + \beta(1-s)} L_f. \label{E:moneyLE}
\end{eqnarray}
For the respective planned jelly demand $J_E$ then
\begin{equation}
	J_E = \frac{w}{p} \cdot \frac{\beta (1-s)}{\alpha + \beta(1-s)} L_f =
	\frac{w}{p} L_E. \label{E:moneyJE}
\end{equation} 
Setting $J_E=L_E^\gamma$ yields for the (temporarily stable) economic
equilibrium wage-to-price ratio
\begin{equation}
 	\left( \frac{w}{p} \right)_E = \left( \frac{\beta
 	(1-s)}{\alpha+\beta(1-s)} L_f \right)^{\gamma-1}. \label{E:moneyWpE}
\end{equation}
Depending on $w, p, M$ there might be other economic equilibria if
$J_H=L_H^\gamma$ but (\ref{E:moneyWpE}) holds for the ones that are stable
over time.

\subsection{Dynamic Equilibria, Comparison with Previous Cases}

In the extended model, there is one more market and one more price (with the
respective expectation-and-updating routine by the firm) and thus the resulting
dynamics of the system become more complex. A detailed analysis is beyond the
scope of this paper and will be subject of further research. However, some
similarities and differences to the previous cases are pointed out in the following.

A couple of propositions have been made about the dynamics of the original
model. For the extended model, first simulations suggest that the same types of
fixed points can be found for the extended model, so it seems that Proposition \ref{P:types} applies for the extended model as well.

\begin{figure}
\begin{center}
\includegraphics[width=5cm]{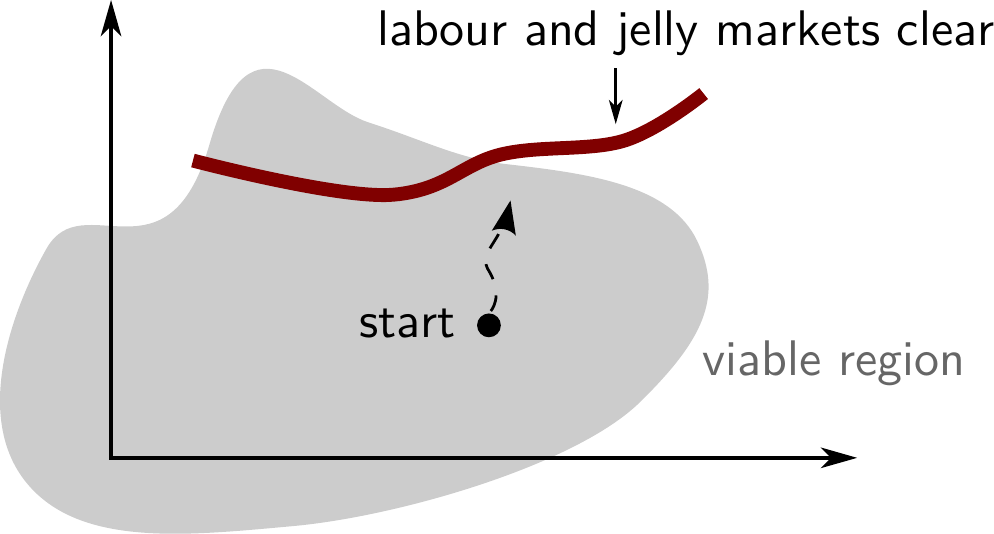}
\caption{With money as a third good, there is more than
one point in the economic space where jelly and labour markets clear. These
points can lie inside, outside or partly in the viable region.
\label{F:stateSpaceMoney}}
\end{center}
\end{figure}

However, there is a fundamental difference between the extended model and the
previous cases: equations (\ref{E:LE})-(\ref{E:wpE}) determine a single point in
the economic space but equations (\ref{E:moneyLE})-(\ref{E:moneyWpE}) an
infinite set of points, i.e. for the previous cases the economic equilibrium is
a single point in the economic space but for the extended model it is a larger
set. The set of economic equilibria can also lie partly in the viable region,
as depicted in Figure \ref{F:stateSpaceMoney}.

This can be illustrated e.g. reformulating Proposition \ref{P:minMoney} for the
extended model. This proposition states that (for the previous case with money)
it is a necessary condition for the economic equilibrium that there is a minimum
total amount of money $M=L_E$ in the system. For the extended model, the
respective minimum amount of money would be $M=(1+s)w\cdot L_E$ (there has to
be enough money for paying the equilibrium income plus the amount the household
keeps under the mattress), i.e. it depends on the wage level, whether the
necessary condition for the total amount of money is fulfilled or not.

Of course, here we defined the economic equilibrium only by clearing labour
and jelly markets while not paying attention to the money market. The full
analysis of the three goods case (jelly, labour, money) is beyond the scope of
this paper, and will be pursued in future work. But the findings of this paper
already suggest that this analysis will focus on the question what kind of
monetary policies make economic equilibria viable.

%=======================================================================================

\section{Conclusions}
\label{S:conclusions}

It is an open question in economic theory how to extend the mathematical model
of general equilibrium theory to include price adjustments. Removing the
fictitious auctioneer from the canonical Arrow-Debreu framework raises the
question of how prices are set and adjusted by decentralised actors with incomplete
information. 

In this paper, we considered a very basic model with only two aggregate agents,
a household and a firm. The firm sets the prices. It compensates for the lack of
information about the household's demand using expectations about that. It
improves its expectations about the household's demand by observing the
household's reactions to past prices.

We investigated the dynamics of this system with respect to the
question whether the economic equilibrium is obtained eventually. We found
that it is essential for such a system that value can be stored over time. For
this purpose, it seems to be natural to introduce money to the modelled economy,
but this raises questions about whether and to which degree money is a
commodity of its own, and how to determine supply and demand for it.

One main point of this paper is that we look at the trajectories of the system
in what we call the economic space, i.e. a projection of the state space onto
the ``economic'' state variables: supplies, demands, and prices. While the
system is deterministic, evolution in the economic space is not (i.e. only
knowing the ``economic'' variables does not suffice to predict the future
development). But statements can be made about which parts of the economic space
are allowed for evolution, and the system's ``freedom'' to evolve over time is
strongly to connected to the availability of money to the agents. The task to
model an out-of-equilibrium price dynamics based on expectations brought up the
need for money in the system ``naturally''. Therefore, we consider this paper as
a first step towards modelling an out-of-equilibrium economy with money and
(at least) two other commodities.

%=======================================================================================
\begin{acknowledgements}
Various discussions with Andreas Geiges and Franziska Sch\"{u}tze have
been very useful and are highly acknowledged, as well as valuable programming
support by Steffen F\"{u}rst.
\end{acknowledgements}

\section*{Appendix: The Dynamic System}

In the following, the dynamic system is formulated for all three versions of the
model. 

\begin{enumerate}

\item \textbf{Firm decides production}\\
Versions 1 and 2, labour \num:
\begin{eqnarray}
L^{(D)}_{t+1} &=& \mathrm{arg}\min_L \vert \rho(L)\cdot
	\phi_t(\rho(L)) - L \vert \\
& & \text{with} \;\; \rho(L)= L^\gamma\\
& & \text{and} \;\; \phi_t(\rho(L))=
	\frac{z_t}{\rho(L)^{\zeta_t}}\\
	& & \text{and} \;\; 0\le L \le L_f\\
  & & \text{and} \;\; L\le \left\{ \begin{array}{ll} 
		 p_t \cdot J^{(M)}_t& \; (\text{version}\; 1)\\ 
		 m^{(F)}_t & \; (\text{version}\; 2) 
  \end{array} \right.\\
J^{(S,\text{planned})}_{t+1} &=& \rho\left(L^{(D)}_{t+1}\right) =
	\left(L^{(D)}_{t+1}\right)^{\gamma}\\
p_{t+1} &=& \phi_t\left(J^{(S,\text{planned})}_{t+1}\right)
\end{eqnarray} 
Version 3,  money \num:
\begin{eqnarray}
L^{(D)}_{t+1} &=& \mathrm{arg}\min_L \vert \rho(L) \cdot
	\phi_t(\rho(L)) - \psi_t(L)\cdot L\vert \\
& & \text{with} \;\; \rho(L)= L^\gamma\\
& & \text{and} \;\; \phi_t(\rho(L))= \frac{z_t}{\rho(L)^{\zeta_t}}\\
& & \text{and} \;\; \psi_t(L)=
	\left(\frac{x_t}{L_f-L}\right)^\frac{1}{\xi_t}\\
& & \text{and} \;\; 0\le L \le L_f\\
& & \text{and} \;\; L \le \mathrm{arg}\min_{L\ge0} \vert
	\psi_t(L) - \frac{m^{(F)}_t}{L}\vert \\
J^{(S,\text{planned})}_{t+1} &=& \rho\left(L^{(D)}_{t+1}\right) =
	\left(L^{(D)}_{t+1}\right)^{\gamma}\\
p_{t+1} &=& \phi_t\left(J^{(S,\text{planned})}_{t+1}\right)\\
w_{t+1} &=& \psi_t\left(L^{(D)}_{t+1}\right) = 
	\left(\frac{x_t}{L_f-L^{(D)}_{t+1}}\right)^\frac{1}{\xi_t}
\end{eqnarray}

\item \textbf{Household decides labour}\\
Versions 1 and 2, labour \num:
\begin{equation}
L^{(S)}_{t+1} = L_f \cdot \frac{\beta}{\alpha +\beta} 
\end{equation}
Version 3,  money \num:
\begin{equation}
L^{(S)}_{t+1} = L_f \cdot \frac{\beta}{\alpha+\beta} -
\frac{\alpha}{\alpha+\beta}\cdot
	\frac{m^{(H)}_t}{(1-s)\cdot w_{t+1}}
\end{equation}

\item \textbf{Labour market transaction} (all versions)
\begin{equation}
L^{(M)}_{t+1} = \min \left\{ L^{(D)}_{t+1},L^{(S)}_{t+1} \right\}
\end{equation}

\item \textbf{Firm produces jelly} (all versions)
\begin{eqnarray}
J^{(S)}_{t+1} &=& (L^{(M)}_{t+1})^\gamma
\end{eqnarray}

\item \textbf{Household decides consumption}\\
Version 1, labour \num:
\begin{equation}
J^{(D)}_{t+1} = \frac{L^{(M)}_{t+1}}{p_{t+1}}
\end{equation}
Versions 2, labour \num:
\begin{equation}
J^{(D)}_{t+1} = \min \left\{ \frac{\beta}{\alpha+\beta} \cdot
\frac{L_f}{p_{t+1}} , \frac{m^{(H)}_t}{p_{t+1}} \right\}
\end{equation}
Version 3,  money \num:
\begin{equation}
J^{(D)}_{t+1} = \min \left\{ \frac{\beta}{\alpha+\beta} \cdot \left( (1-s)
\frac{w_{t+1}}{p_{t+1}} L_f + \frac{m^{(H)}_t}{p_{t+1}} \right),
\frac{m^{(H)}_t}{p_{t+1}} \right\}
\end{equation}

\item \textbf{Jelly market transaction} (all versions)
\begin{equation}
J^{(M)}_{t+1} = \min \left\{ J^{(D)}_{t+1},J^{(S)}_{t+1} \right\}
\end{equation}

\item \textbf{Firm updates monetary holdings} (only versions 2 and 3)
\begin{equation}
m^{(F)}_{t+1} = m^{(F)}_t + p_{t+1}\cdot J^{(M)}_{t+1}-w_{t+1}\cdot
L^{(M)}_{t+1}
\end{equation}
(with $w_{t+1}=1$ in version 2)\\

\item \textbf{Firm updates expected jelly-demand function} (all versions)
\begin{eqnarray}
(\zeta_{t+1},z_{t+1}) &=&
	\mathrm{arg}\min_{\zeta, z}
	\left(\sqrt{\frac{\ln^2\Delta^{(p)}_{t+1}+\epsilon_1 \cdot
	\ln^2\Delta^{(p)}_{t}} {1+\epsilon_1}} + \epsilon_2
 	\sqrt{\ln^2\frac{\zeta}{\zeta_t}+\ln^2\frac{z}{z_t}} \right)\nonumber\\
 	&&  \text{with} \;\; \Delta^{(p)}_\tau =
 	\frac{p_\tau}{\phi(J^{(D)}_\tau)}\nonumber\\
 	&&  \text{and} \;\; z>0, \; 1>\zeta>0 \nonumber
\end{eqnarray}

\item \textbf{Firm updates expected labour-supply function} (only version 3)
\begin{eqnarray}
(\xi_{t+1},x_{t+1}) &=& \mathrm{arg}\min_{\xi,x}
	\left(\sqrt{\frac{\ln^2\Delta^{(w)}_{t+1}+\epsilon_1 \cdot
	\ln^2\Delta^{(w)}_{t}} {1+\epsilon_1}} + \epsilon_2
	\sqrt{\ln^2\frac{\xi}{\xi_t} + \ln^2\frac{x}{x_t}+} \right)\\
	& & \text{with} \;\; \Delta^{(w)}_\tau =
	\frac{w_\tau}{\psi(L^{(S)}_\tau)}\\
	& & \text{and} \;\; x > 0, \; \xi > 0
\end{eqnarray}

\item \textbf{Household updates monetary holdings} (only versions 2 and 3)
\begin{equation}
m^{(H)}_{t+1} = m^{(H)}_t - p_{t+1}\cdot J^{(M)}_{t+1} + w_{t+1}\cdot
L^{(M)}_{t+1}
\end{equation}
(with $w_{t+1}=1$ in version 2)\\

\end{enumerate}

\end{document}